\documentclass[pra,twocolumn,twoside,showpacs]{revtex4}

\usepackage{graphicx}
\usepackage{amsmath,amssymb,bm}

\newcommand{\eq}[1]{\begin{equation}#1\end{equation}}
\newcommand{\eqmulti}[1]{\begin{equation}\begin{split}#1\end{split}\end{equation}}
\newcommand{\eqgather}[1]{\begin{gather}#1\end{gather}}
\newcommand{\eqalignat}[2]{\begin{alignat}{#1}#2\end{alignat}}

\newcommand{\bra}[1]{\ensuremath{ \big< {#1} \big| \, }}
\newcommand{\ket}[1]{\ensuremath{ \, \big| {#1} \big> }}
\newcommand{\braket}[2]{\ensuremath{ \big< {#1} \big| {#2} \big> }}
\newcommand{\matrixe}[3]{\ensuremath{ \big< {#1} \big| \,{#2}\, \big| {#3} \big> }}
\newcommand{\expect}[1]{\ensuremath{ \big< #1 \big> }}
\newcommand{\op}[1]{\ensuremath{\bm{#1}}}
\newcommand{\dd}{\ensuremath{\mathrm{d}}}
\newcommand{\ii}{\ensuremath{\mathrm{i}}}

\newcommand{\qO}{\ensuremath{\op{q}}}
\newcommand{\rO}{\ensuremath{\op{r}}}
\newcommand{\vO}{\ensuremath{\op{v}}}

\newcommand{\HO}{\ensuremath{\op{H}}}
\newcommand{\UO}{\ensuremath{\op{U}}}

\newcommand{\PiO}{\ensuremath{\op{\Pi}}}

\newcommand{\kV}{\ensuremath{\vec{k}}}

\newcommand{\rV}{\ensuremath{\vec{r}}}
\newcommand{\xV}{\ensuremath{\vec{x}}}

\newcommand{\pOV}{\ensuremath{\vec{\op{p}}}}
\newcommand{\qOV}{\ensuremath{\vec{\op{q}}}}
\newcommand{\rOV}{\ensuremath{\vec{\op{r}}}}
\newcommand{\rUOV}{\ensuremath{\hat{\op{r}}}}
\newcommand{\xOV}{\ensuremath{\vec{\op{x}}}}

\newcommand{\EC}{\ensuremath{\mathcal{E}}}
\newcommand{\FC}{\ensuremath{\mathcal{F}}}

\newcommand{\eff}{\ensuremath{\text{eff}}}
\newcommand{\exact}{\ensuremath{\text{exact}}}
\newcommand{\pseudo}{\ensuremath{\text{pseudo}}}

\allowdisplaybreaks[3]

\begin{document}

\title{Effective s- and p-Wave Contact Interactions\\
  in Trapped Degenerate Fermi Gases}

\author{R. Roth}
\email{r.roth@gsi.de}
\homepage[\\ \indent \hskip0.8em URL: ]{http://theory.gsi.de/~trap/}

\author{H. Feldmeier}
\email{h.feldmeier@gsi.de}

\affiliation{Gesellschaft f\"ur Schwerionenforschung (GSI), 
  Planckstr. 1, 64291 Darmstadt, Germany}

\date{\today}

\begin{abstract}
\vskip5pt 

The structure and stability of dilute degenerate Fermi gases trapped
in an external potential is discussed with special emphasis on the
influence of s- and p-wave interactions. In a first step an Effective
Contact Interaction for all partial waves is derived, which reproduces
the energy spectrum of the full potential within a mean-field model
space. Using the s- and p-wave part the energy density of the
multi-component Fermi gas is calculated in Thomas-Fermi
approximation. On this basis the stability of the one- and
two-component Fermi gas against mean-field induced collapse is
investigated. Explicit stability conditions in terms of density and
total particle number are given. For the single-component system
attractive p-wave interactions limit the density of the gas. In the
two-component case a subtle competition of s- and p-wave interactions
occurs and gives rise to a rich variety of phenomena. A repulsive
p-wave part, for example, can stabilize a two-component system that
would otherwise collapse due to an attractive s-wave interaction. It
is concluded that the p-wave interaction may have important influence
on the structure of degenerate Fermi gases and should not be discarded
from the outset.

\end{abstract}


\maketitle



\section{Introduction}

The achievement of Bose-Einstein condensation in trapped dilute gases
of bosonic atoms \cite{AnEn95} triggered a wide spread interest in
the field of ultracold atomic gases. Meanwhile these systems appear as
a unique lab for the study of all kinds of fundamental quantum
phenomena. 

After a series of excellent experiments on bosonic systems the
question arises, whether a dilute gas of fermionic atoms can also be
cooled to temperatures where quantum effects dominate. In 1999 the
group of Deborah Jin managed to cool a sample of typically $10^6$
fermionic ${}^{40}$K atoms to temperatures significantly below the
Fermi energy of the system \cite{DeJi99}; in recent experiments they
achieved a temperature corresponding to one fifth of the Fermi energy
\cite{DePa01}. In this temperature regime the system can be described
as a degenerate Fermi gas, where the majority of the atoms
successively fills the lowest available one-body states according to
the Pauli principle.

One of the goals of the investigations on dilute and ultracold Fermi
gases is the observation of Cooper pairing and the transition to a
superfluid state. The transition temperature depends on the density
and on the strength of the attractive interaction that is necessary
for the formation of Cooper pairs \cite{HoFe97,StHo98}. In order to
increase the transition temperature one may increase the density or
the interaction strength, where the latter seems to be more
promising. An atomic species favored for the experimental observation
of a BCS transition is ${}^{6}$Li due to its large s-wave scattering
length of $a_0\approx-2160a_{\text{B}}$ \cite{AbMc97}. But also
${}^{40}$K, which shows a rather small natural scattering length, is a
possible candidate for Cooper pair formation \cite{DePa01}, because a
simultaneous s- and p-wave Feshbach resonance is predicted
\cite{Bohn00}, which allows tuning of the s- and p-wave scattering
lengths over a very wide range.

A serious constraint on the way towards a superfluid Fermi gas is the
mechanical stability of the progenitor, i.e., the normal degenerate
Fermi gas.  As was experimentally demonstrated for the bosonic
${}^{85}$Rb system \cite{CoCl00}, an attractive interaction between
the atoms leads to a mean-field instability of the trapped gas if
the density exceeds a critical value. A similar collapse is expected in
fermionic systems with attractive interactions. In contrast to the
bosonic systems p-wave interactions contribute in a Fermi gas and may
have strong influence on the stability of the system \cite{RoFe00b}.

In the following we address the question of the stability of
degenerate one- and two-component Fermi gases in presence of s- and
p-wave interactions within a simple and transparent model. In Section
\ref{sec:eci} we derive an Effective Contact Interaction (ECI) for all
partial waves that is suited for a mean-field treatment of the
many-body problem. Using the s- and p-wave part of this interaction we
construct in Section \ref{sec:ef} the energy functional of a trapped
multi-component Fermi gas in Thomas-Fermi approximation. From that the
ground state density profile can be determined by functional
variation. In Section \ref{sec:1comp} we discuss the structure and
stability of single-component Fermi gases, where only the p-wave
interaction contributes according to the Pauli principle. Section
\ref{sec:2comp} deals with two-component Fermi gases where s- and
p-wave interactions are present and lead to a subtle dependence of the
stability on the two scattering lengths.

\section{Effective Contact Interaction}
\label{sec:eci}

\subsection{Concept}
\label{sec:eci_concept}

The approximate solution of the many-body problem in a restricted
low-momentum sub-space of the full Hilbert space faces a fundamental
problem: Many realistic two-body interactions --- like
van-der-Waals-type interactions between atoms or the interactions
between two nucleons in an atomic nucleus --- exhibit a strong
short-range repulsion. This repulsion generates particular short-range
correlations in the many-body state, which cannot be described within
a low-momentum model-space \cite{UCOM98}. Therefore short-range
correlations inhibit the use of a realistic two-body interactions in
the framework of a naive mean-field model. One way to overcome this
problem is to replace the original interaction by a suitable effective
interaction. How this effective interaction has to be constructed from
the original potential depends on the properties of the actual
physical system under investigation.

Cold and very dilute quantum gases allow an effective interaction of
simple structure. The typical wave length of the relative motion of
two particles is always large compared to the range of the
interaction. Therefore the particles experience only an ``averaged''
two-body potential and do not probe the detailed radial dependence.
Moreover the gases are in a metastable not self-bound state that is
kept together by the external trapping potential. Thus the bound
states of the two-body potential are not populated and have only indirect
influence. We make use of these facts and replace the original
potential by a contact interaction. The strength of the contact terms
is related to the properties of the original potential.

\begin{figure}
\includegraphics[height=0.27\textheight]{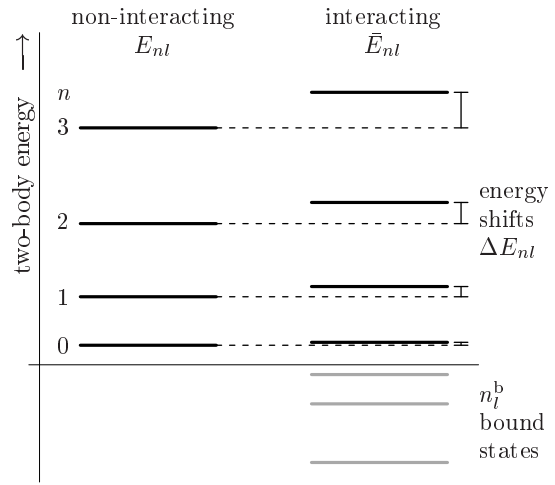}
\caption{Schematic comparison of the free two-body energy spectrum $E_{nl}$
  with the energy spectrum $\bar{E}_{nl}$ in presence of a two-body
  interaction for given relative angular momentum $l$. The interaction
  generates $n^{\text{b}}_l$ bound states and shifts the positive
  energy states by $\Delta E_{nl}$.}
\label{fig:eci_spectrumschematic}
\end{figure}

We establish this relation via the energy spectrum of the two-body
system. Consider a two-body problem with an auxiliary boundary
condition at some large radius such that the spectrum is discrete even
for positive energies. For noninteracting particles this spectrum
shows a sequence of levels at positive energies $E_{nl}$ which are
labeled by an angular momentum quantum number $l$ and a radial quantum
number $n$. A schematic sketch of this spectrum for some fixed $l$ is
shown in Figure \ref{fig:eci_spectrumschematic}. If we switch on an
interaction between the particles two things happen: A number
$n^{\text{b}}_l$ of bound two-body states with relative angular
momentum $l$ may appear at negative energies and the levels at
positive energies $\bar{E}_{nl}$ are shifted compared to the
noninteracting spectrum. For our purpose this energy-shift $\Delta
E_{nl} = \bar{E}_{nl} - E_{nl}$ is defined between the positive energy
states only. The lowest level of positive energy is labeled with the
quantum number $n=0$.

The energy-shift is the relevant ``averaged'' property of the
two-body potential that the effective interaction should reproduce. We
require that the expectation value of the effective interaction
calculated with eigenstates $\ket{nlm}$ of the noninteracting
two-body problem equal the energy shift $\Delta E_{nl}$ induced by the
original potential 
\eq{ \label{eq:eci_energyshiftcondition}
  \matrixe{nlm}{\vO^{\eff}}{nlm} 
  \overset{!}{=} \bar{E}_{nl} - E_{nl} = \Delta E_{nl} .
}
Due to the angular momentum dependence of the energy-shifts we have to
construct the effective contact interaction as a sum of contact
interaction for each partial wave. The strength of each contact term
is fixed by this condition.

\subsection{Construction of the ECI}
\label{sec:eci_derivation}

The construction of the Effective Contact Interaction (ECI) is
organized in two steps: Firstly, the two-body energy-shift induced by
the original potential is evaluated. Secondly, the operator of the
Effective Contact Interaction is formulated for each partial wave and
the interaction strengths are fixed by condition
\eqref{eq:eci_energyshiftcondition}.

\subsubsection{Energy Shift}

Assume a system of two particles with reduced mass $\mu=m_1
m_2/(m_1+m_2)$. Their wave function can be decomposed into a center of
mass and a relative wave function, where the latter separates into a
radial and a angular component because of rotational symmetry
\eq{
  \braket{\rV}{nlm} = R_{nl}(r)\, Y_{lm}(\Omega) .
}
The radial wave functions  $R_{nl}(r)$ of the noninteracting two-body
system are solutions of the Schr\"odinger equation
\eq{ \label{eq:eci_radialschroed_nonint}
  \bigg[\!-\!\frac{1}{r} \frac{\partial^2}{\partial r^2} r 
  + \frac{l(l+1)}{r^2} - 2\mu E_{nl} \bigg] R_{nl}(r) = 0 
}
(we use units with $\hbar=1$). We require the radial wave function to
vanish at some arbitrary but large radius $\Lambda$
\eq{ \label{eq:eci_boundarycond}
  R_{nl}(\Lambda) \overset{!}{=} 0 .
}
This auxiliary boundary condition leads to a discrete energy spectrum,
which is needed to enumerate the energy levels and to evaluate the
energy shift.

In the noninteracting case the solution of the radial Schr\"odinger
equation \eqref{eq:eci_radialschroed_nonint} is given by a spherical Bessel
function $j_l(x)$
\eq{ \label{eq:eci_wavefunc_nonint}
  R_{nl}(r) = A_{nl}\, j_l(q_{nl}r) ,
}
where $q_{nl}$ denotes the relative momentum of the two particles and
$A_{nl}$ a normalization constant. The discrete momenta $q_{nl}$ are
determined by the boundary condition \eqref{eq:eci_boundarycond} and
thus are related the to the zeros of the Bessel function. Since the radius
$\Lambda$ can be chosen arbitrary large it is sufficient to use the
asymptotic expansions of the spherical Bessel and Neumann functions
\eqmulti{ \label{eq:eci_besselasymp}
  j_l(x) &\overset{x\gg l}{=} 
    \phantom{-}\frac{1}{x} \sin\big(x-\pi l/2\big) , \\
  n_l(x) &\overset{x\gg l}{=} 
    -\frac{1}{x} \cos\big(x-\pi l/2\big) .
}
Evaluating the boundary condition \eqref{eq:eci_boundarycond} with the
asymptotic form of the Bessel function we obtain for the possible
relative momenta
\eq{ \label{eq:eci_momentum_nonint}
  q_{nl}\Lambda = \pi(n + l/2) .
} 
Accordingly the two-body energy spectrum in the noninteracting case
is given by
\eq{
  E_{nl} 
  = \frac{1}{2\mu} q_{nl}^2 
  = \frac{\pi^2}{2\mu\Lambda^2} (n +l/2)^2 .
}
The normalization constant can be determined explicitly
\eq{
  A_{nl}^{-2} 
  = \int_0^{\Lambda}\!\!\dd r\;r^2 j_l^2(q_{nl}r)
  = \frac{\Lambda^3}{2} j_{l+1}^2(q_{nl}\Lambda) .
}
Inserting the asymptotic form of the Bessel function we finally get
\eq{ \label{eq:eci_normconstant}
  A_{nl}^{-2} 
  \overset{q_{nl}\Lambda \gg l}{=} \frac{\Lambda}{2 q_{nl}^2} .
}

In presence of a two-body potential $v(r)$ of finite range $\lambda$
the solution $\bar{R}_{nl}(r)$ of Schr\"odinger equation
\eq{ \label{eq:eci_radialschroed_int}
  \bigg[\!-\!\frac{1}{r} \frac{\partial^2}{\partial r^2} r 
  + \frac{l(l+1)}{r^2} + 2\mu\big[ v(r)-\bar{E}_{nl} \big] \bigg] 
  \bar{R}_{nl}(r) = 0 
}
outside the range of the potential ($r>\lambda$) is given by a
superposition of spherical Bessel and Neumann functions
\eq{
  \bar{R}_{nl}(r) = \bar{A}_{nl} \big[ j_l(\bar{q}_{nl}r)  
  - \tan\eta_l(\bar{q}_{nl})\; n_l(\bar{q}_{nl}r) \big] ,
}
where $\eta_l(q)$ denotes the phase shift of the potential for the
$l$-th partial wave. The bar distinguishes quantities in presence of
the interaction from those in the noninteracting case.  Imposing the
boundary condition \eqref{eq:eci_boundarycond} we get the following
implicit equation for the momenta $\bar{q}_{nl}$ in the interacting
case
\eq{
  \frac{j_l(\bar{q}_{nl}\Lambda)}{n_l(\bar{q}_{nl}\Lambda)}
  = \tan\eta_l(\bar{q}_{nl}) .
}
Expressing the Bessel and Neumann functions by their asymptotic
expansion \eqref{eq:eci_besselasymp} this reduces to
\eq{
  -\tan( \bar{q}_{nl}\Lambda - \pi l/2 ) 
  = \tan\eta_l(\bar{q}_{nl}) .
}
In order to associate the energy levels with same quantum number $n$
in the interacting and noninteracting case as described above, the
lowest positive-energy state should be labeled with the quantum number
$n=0$.  To achieve this for a potential with $n^{\text{b}}_l$ bound
two-body states with angular momentum quantum number $l$ we add the
phase $\pi(n + n^{\text{b}}_l)$ to the argument on the r.h.s., i.e.,
the bound states contribute according to Levinson's theorem. The
momenta $\bar{q}_{nl}$ in presence of the interaction are thus
determined by the equation
\eq{ \label{eq:eci_momentum_int}
  \bar{q}_{nl}\Lambda 
  = -\eta_l(\bar{q}_{nl}) + \pi(n + n^{\text{b}}_l + l/2) .
}

The momentum shift $\Delta q_{nl}$ induced by the interaction is
obtained by subtracting the momenta of the noninteracting system
\eqref{eq:eci_momentum_nonint} from equation
\eqref{eq:eci_momentum_int}
\eqmulti{ \label{eq:eci_momentumshift_1}
  \Delta q_{nl} \Lambda
  &= (\bar{q}_{nl} - q_{nl}) \Lambda \\
  &= -[\eta_l(\bar{q}_{nl}) - \pi n^{\text{b}}_l] 
  = -\tilde{\eta}_l(\bar{q}_{nl}) .
}
Here $\tilde{\eta}_l(q) = \eta_l(q) - \pi n^{\text{b}}_l$ denotes the
phase shift reduced by the contribution of the bound states. In a
final step we expand the phase shifts around the momenta $q_{nl}$ of
the noninteracting system
\eq{
  \tilde{\eta}_l(\bar{q}_{nl}) 
  = \tilde{\eta}_l(q_{nl}) + \tilde{\eta}'_l(q_{nl}) \Delta q_{nl} 
  + \cdots .
}
Already the term linear in $\Delta q_{nl}$ can be neglected in good
approximation because the momentum shift is of the order $1/\Lambda$
according to equation \eqref{eq:eci_momentumshift_1}. Thus we retain
the following simple expression for the momentum shift
\eq{
  \Delta q_{nl}\Lambda = - \tilde{\eta}_l(q_{nl}) .
}

The shift of the energy levels of the interacting two-body system
compared to the noninteracting spectrum is connected with the momentum
shift by
\eq{
  \frac{\Delta E_{nl}}{E_{nl}} 
  = 2 \frac{\Delta q_{nl}}{q_{nl}} + \bigg(\frac{\Delta
  q_{nl}}{q_{nl}}\bigg)^2 .
}  
The term quadratic in the momentum shift can be neglected because
$\Delta q_{nl}/q_{nl}$ is small. The final expression for the energy shift
reads
\eq{ \label{eq:eci_energyshift}
  \frac{\Delta E_{nl}}{E_{nl}} 
  = -\frac{2}{\Lambda}\; \frac{\tilde{\eta}_l(q_{nl})}{q_{nl}} .
}
The proportionality between energy shift and phase shifts is well
known \cite{Gott66} and was used in different applications before.

\subsubsection{ECI in Scattering Length Approximation}

In a second step we construct an operator form of the Effective
Contact Interaction that obeys condition
\eqref{eq:eci_energyshiftcondition}. For the application to ultracold
dilute quantum gases the general form \eqref{eq:eci_energyshift} of
the energy shift $\Delta E_{nl}$ can be simplified considerably. The
relative momenta in these systems are extremely low, i.e., typical
relative wave lengths are large compared to the range of the
interaction. This allows an expansion of the phase shifts
$\tilde{\eta}_l(q)$ in a power series in relative momentum. In lowest
order approximation ($q\,a_l\ll 1$) the phase shifts of the $l$-th
partial wave can be expressed in terms of the corresponding scattering
length $a_l$ \footnote{Some authors \cite{Bohn00} use a different
definition of the p-wave scattering length without the $l$-dependent
prefactors. We use this general form (see \cite{HuYa57}) where the
scattering length for a hard sphere equals the sphere radius for all
$l$.}
\eq{ \label{eq:eci_scatteringlength}
  \frac{\tilde{\eta}_l(q)}{q^{2l+1}} \approx
  \frac{\tan\eta_l(q)}{q^{2l+1}} \approx -\frac{(2l+1)}{[(2l+1)!!]^2}
  \; a_l^{2l+1} .  
}
The energy shift in this scattering length approximation is given by
\eq{ \label{eq:eci_energyshift_scattapprox}
  \frac{\Delta E_{nl}}{E_{nl}} 
  = \frac{2}{\Lambda}\,\frac{(2l+1)}{[(2l+1)!!]^2}\;
    q_{nl}^{2l}\, a_l^{2l+1} .
}
Based on this form we can construct a simple operator for the ECI. For
applications where the approximation \eqref{eq:eci_scatteringlength}
is not sufficient higher order terms of the power series can be
included successively. We will come back to this point in the
following section.
 
According to the dependence of the energy shifts on the angular
momentum quantum number $l$ the operator of the effective interaction
$\vO^{\eff}$ is formulated as a sum of independent operators
$\vO^{\eff}_l$ for each partial wave
\eq{
  \vO^{\eff} = \sum_{l=0}^{\infty} \PiO_l\, \vO^{\eff}_l\, \PiO_l ,
}
where $\PiO_l$ denotes the projection operator on the subspace spanned
by states of relative angular momentum $l$. We want to use a contact
interaction for each partial wave. This requires nonlocal interaction
operators beyond $l=0$, i.e., derivative couplings. The simplest
ansatz for the operator of the effective contact interaction for the
$l$-th partial wave is
\eqmulti{ \label{eq:eci_interactionoperator}
  \vO^{\eff}_l 
  &= (\qOV\cdot\rUOV)^l\; 
    g_l\, \delta^{(3)}(\rOV)\; (\rUOV\!\cdot\qOV)^l \\
  &= \int\!\!\dd^3r \ket{\rV}
    \frac{\overset{\leftarrowtail}{\partial}{}^l}{\partial r^l}\; 
    g_l\, \delta^{(3)}(\rV)\; 
    \frac{\overset{\rightarrowtail}{\partial}{}^l}{\partial r^l}\bra{\rV}.
}
Here $\qOV = \tfrac{1}{2}(\pOV_1 - \pOV_2)$ denotes the operator of
the relative momentum of two particles, $\rUOV = \rOV/\rO$ the unit
vector along the relative coordinate and $\delta^{(3)}(\rV) =
\tfrac{1}{4\pi r^2}\delta(r)$ the radial component of the
3-dimensional delta function. The arrows above the derivatives
indicate to which side they act. 

The interaction strength $g_l$ is a constant that contains the
relevant information on the original two-body potential.  The
connection is provided by condition
\eqref{eq:eci_energyshiftcondition} via the shift of the energy levels
with respect to the free spectrum. To evaluate
\eqref{eq:eci_energyshiftcondition} we calculate the expectation value
of the contact interaction \eqref{eq:eci_interactionoperator} for the
$l$-th partial wave taken with the noninteracting two-body states
$\ket{nlm}$
\eqmulti{ \label{eq:eci_expectationvalue}
  &\matrixe{nlm}{\vO^{\eff}_l}{nlm} \\
  &\qquad= g_l \int\!\!\dd^3r\; \delta^{(3)}(\rV)\, 
    \Big| \frac{\partial^l}{\partial r^l} R_{nl}(r)  Y_{lm}(\Omega) \Big|^2 \\
  &\qquad= \frac{g_l}{4\pi}\; \Big| \frac{\partial^l}{\partial r^l} 
    R_{nl}(r) \Big|^2_{r=0} \\
  &\qquad= \frac{g_l}{4\pi}\;\bigg[\frac{l!}{(2l+1)!!}\bigg]^2 
    A_{nl}^2\, q_{nl}^{2l} , 
}
where we used the expansion of the radial wave-function around the origin
\eq{
  R_{nl}(r) = A_{nl} \frac{(q_{nl} r)^l}{(2l+1)!!}
    \bigg[ 1 - \frac{(q_{nl}r)^2}{2(2l+3)} + \cdots \bigg] .
}
Equating the expectation value \eqref{eq:eci_expectationvalue} with
the energy shift results in the following equation for the interaction
strengths
\eq{
  g_l 
  = 4\pi \bigg[\frac{(2l+1)!!}{l!}\bigg]^2 A_{nl}^{-2}
    \frac{\Delta E_{nl}}{q_{nl}^{2l}} 
}
Inserting the normalization constant \eqref{eq:eci_normconstant} and the
energy shift in scattering length formulation
\eqref{eq:eci_energyshift_scattapprox} gives the final
expression for the interaction strengths
\eq{ \label{eq:eci_interactionstrength_scattapprox}
  g_l
  = \frac{4\pi}{2\mu}\, \frac{(2l+1)}{(l!)^2}\; a_l^{2l+1} .
}
Together with equation \eqref{eq:eci_interactionoperator} this defines
the scattering length formulation of the Effective Contact
Interaction.  Notice that these expressions do not depend on the
auxiliary boundary condition \eqref{eq:eci_boundarycond}, which was
introduced to obtain a discrete energy spectrum.

For $l=1$ sometimes a gradient operator with respect to the relative
coordinate is used instead of the radial derivative
\eqref{eq:eci_interactionoperator}. This alternative ansatz reads
\eqmulti{ \label{eq:eci_interactionoperator_pwave}
  \vO^{\eff}_1 
  &= \qOV\;\,\tilde{g}_1\,\delta^{(3)}(\rOV)\;\, \qOV \\[2pt]
  &= \int\!\!\dd^3r \ket{\rV}
    \overset{\leftarrowtail}{\nabla}\; 
    \tilde{g}_1\, \delta^{(3)}(\rV)\; 
    \overset{\rightarrowtail}{\nabla}\bra{\rV}.
}
It should be noted that the p-wave interaction strength $\tilde{g}_1$
in the gradient formulation is related to the interaction strength
\eqref{eq:eci_interactionstrength_scattapprox} by
\eq{
  \tilde{g}_1 = g_1/3 .
}

\subsubsection{Beyond Scattering Length Approximation}
\label{sec:eci_general_form}

The operator of the Effective Contact Interaction can be generalized
systematically beyond the scattering length approximation shown in the
preceding section. A formal scheme emerges from the expansion of the
phase shifts in a power series in $q$
\eq{ \label{eq:eci_phaseshift_expansion}
  \frac{\tilde{\eta}_l(q)}{q^{2l+1}} 
  = \sum_{\nu=0}^{\infty} \frac{1}{\nu!}\, c_l^{(\nu)}\,q^{2\nu} .
}
The momentum independent lowest order term of this expansion matches
the scattering length approximation. From equation
\eqref{eq:eci_scatteringlength} the connection between the coefficient
$c_l^{(0)}$ and the the scattering length $a_l$ becomes obvious
\eq{
  c_l^{(0)} = -\frac{(2l+1)}{[(2l+1)!!]^2} \; a_l^{2l+1} .
}
The coefficient $c_l^{(1)}$ of the quadratic term of the expansion is
connected to the so called effective range or effective volume of the
potential. We will discuss this contribution in more detail later. By
inserting the expansion \eqref{eq:eci_phaseshift_expansion} into the
general formula \eqref{eq:eci_energyshift} for the energy shifts we
obtain
\eq{ \label{eq:eci_energyshift_expansion}
  \frac{\Delta E_{nl}}{E_{nl}} 
  = -\frac{2}{\Lambda}\; \sum_{\nu=0}^{\infty} 
     \frac{1}{\nu!}\,c_l^{(\nu)}\,q_{nl}^{2l+2\nu} .
}

Following the basic concept of the ECI these energy shifts have to be
generated by the operator of the ECI according to condition
\eqref{eq:eci_energyshiftcondition}. To include the momentum
dependence of the energy shifts we have to generalize the ansatz for
the ECI operator compared to the simple momentum-independent
scattering-length formulation \eqref{eq:eci_interactionoperator}. The
further calculation will show that 
\eqmulti{ \label{eq:eci_interactionoperator_expansion}
  \vO^{\eff}_l 
  = \sum_{\nu=0}^{\infty} \frac{1}{2}\,& g_l^{(\nu)}
    \big[ (\qOV\cdot\rUOV)^l\; 
    \delta^{(3)}(\rOV)\; (\rUOV\!\cdot\qOV)^{l+2\nu}\\[-5pt]
  &\;\;\; + (\qOV\cdot\rUOV)^{l+2\nu}\; 
    \delta^{(3)}(\rOV)\; (\rUOV\!\cdot\qOV)^l \big] 
}
is a proper ansatz for the effective interaction operator for the
$l$-th partial wave. Besides the more complex nonlocal structure a set
of interaction strengths $g_l^{(\nu)}$ ($\nu=0,1,\dots$) for each
partial wave is included. They are related to the coefficients
$c_l^{(\nu)}$ and thus correspond to the different powers of the
momentum in equation \eqref{eq:eci_energyshift_expansion}. To employ
condition \eqref{eq:eci_energyshiftcondition} we calculate the
expectation value of $\vO^{\eff}_l$ with noninteracting two-body
states $\ket{nlm}$
\eqmulti{
  &\matrixe{nlm}{\vO^{\eff}_l}{nlm} \\
  &\quad= \,\frac{1}{4\pi}\, \sum_{\nu=0}^{\infty} g_l^{(\nu)} 
    \Big[\frac{\partial^l}{\partial r^l} R_{nl}(r)\Big]_{\!r=0}
    \Big[\frac{\partial^{l+2\nu}}{\partial r^{l+2\nu}}R_{nl}(r)\Big]_{\!r=0} \\
  &\quad= \frac{A_{nl}^2}{4\pi} \sum_{\nu=0}^{\infty} g_l^{(\nu)}
    \frac{l!}{(2l+1)!!} \frac{(-1)^{\nu}
    (l+2\nu)!}{2^{\nu}\nu!(2l+2\nu+1)!!}\, q_{nl}^{2l+2\nu} ,
}
where the full expansion of the noninteracting radial wave function
\eqref{eq:eci_wavefunc_nonint} around $r=0$ was used \cite{AbSt72}
\eq{
  R_{nl}(r) = A_{nl} \sum_{\mu=0}^{\infty} 
    \frac{(-1)^{\mu}}{2^{\mu} \mu! (2l+2\mu+1)!!} (q_{nl}r)^{l+2\mu} .
}
By inserting the expansion of the energy shifts
\eqref{eq:eci_energyshift_expansion} and the expectation value into
condition \eqref{eq:eci_energyshiftcondition} and comparing the
coefficients for the different powers of the momentum $q_{nl}$ we
obtain
\eq{ \label{eq:eci_interactionstrength_expansion}
  g_l^{(\nu)} = (-1)^{\nu+1} \frac{4\pi}{2\mu}  
    \frac{2^{\nu}(2l+1)!!(2l+2\nu+1)!!}{l!(l+2\nu)!}\, c_l^{(\nu)} .
}
Thus the interaction strengths $g_l^{(\nu)}$ of the general operator
form of the ECI \eqref{eq:eci_interactionoperator_expansion} are
proportional to the coefficients $c_l^{(\nu)}$ of the expansion of the
phase shifts \eqref{eq:eci_phaseshift_expansion}. The equations
\eqref{eq:eci_interactionoperator_expansion} and
\eqref{eq:eci_interactionstrength_expansion} define the most general
form of the Effective Contact Interaction.

For the application to dilute degenerate Fermi gases we will use the
ECI up to quadratic terms in the momentum, i.e., we include the
scattering length term of the s- and p-wave part as well as the s-wave
effective range correction. At this point we have to discuss the
connection between the quadratic term of the expansion
\eqref{eq:eci_phaseshift_expansion} and the usual effective range
theory. For the s-wave phase-shifts the effective range expansion
reads
\eq{
  q\,\cot\eta_0(q) 
  \approx -\frac{1}{a_0} + \frac{1}{2} r_0\, q^2 ,
}
where $r_0$ is the effective range of the potential. If we convert
this into an expression for $\tilde{\eta}_0(q)/q$ and expand in $q$ we
obtain
\eq{ \label{eq:eci_effrange_expansion}
  \frac{\tilde{\eta}_0(q)}{q} 
  = -a_0  - b_0\, q^2 + \cdots 
}
with an effective volume $b_0$ that depends on the scattering length
$a_0$ and the effective range $r_0$
\eq{ \label{eq:eci_effectivevolume}
  b_0 
  = \tfrac{1}{2} a_0^2 r_0 - \tfrac{1}{3}a_0^3 .
}
Rather than using this relation we will adjust $b_0$ in order to get
the best representation of the phase shift $\tilde{\eta}_0(q)/q$ with
the truncated expansion \eqref{eq:eci_effrange_expansion}.

Finally, inserting $c_0^{(1)}=-b_0$ into equation
\eqref{eq:eci_interactionstrength_expansion} gives an expression for
the interaction strength of the s-wave effective range term
\eq{ \label{eq:eci_interactionstrength_effvolume}
  g_0^{(1)} 
  = -\frac{12\pi}{2\mu}\,b_0 .
}

\subsection{Example: Square-Well Potential}
\label{sec:eci_example}

To illustrate the concept of the Effective Contact Interaction we use the
simple toy-problem of two particles interacting by an attractive
square-well potential of radius $\lambda$ and depth $-V_0$.

\begin{figure*}
\includegraphics[width=0.98\textwidth]{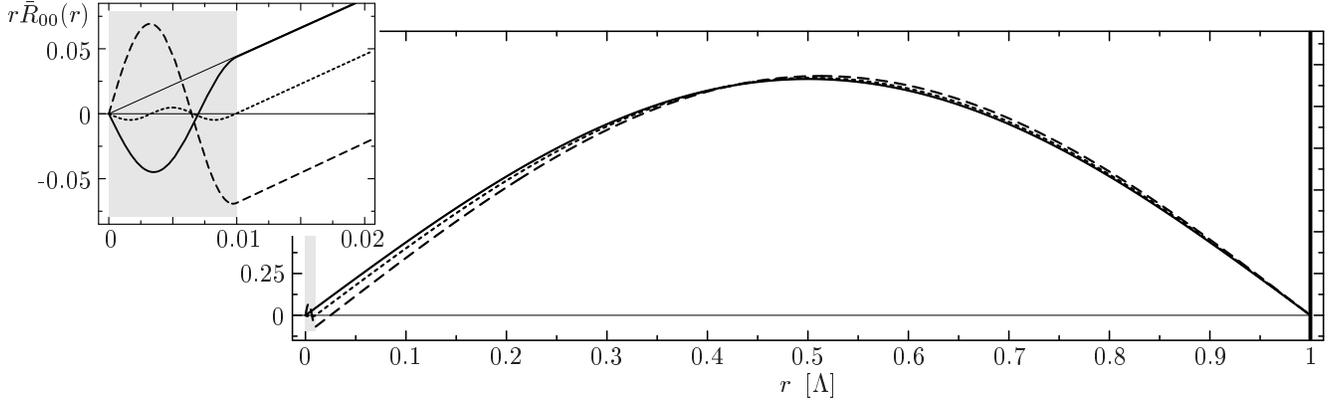}
\caption{Radial wave functions $r \bar{R}_{00}(r)$ of the lowest $l=0$ positive
  energy state for different potential depths $V_0$ of the square-well
  potential. The interaction strengths are $\lambda\sqrt{V_0}=0$ (thin
  solid), $4.49$ (solid), $4.85$ (dashed), and $9.5$ (dotted). The
  radius of the well is $\lambda=0.01\Lambda$ and marked by the gray
  area. The insert shows a magnification of the region around the
  origin.}
\label{fig:eci_rectwell_wavefunc}
\end{figure*}

First we look at typical wave functions to which the idea of the ECI
applies. The major condition is that the typical wavelength of the
relative motion is large compared to the range of the
interaction. This is ensured by choosing the radius $\lambda$ of the
square-well much smaller than the radius $\Lambda$ associated with the
boundary condition \eqref{eq:eci_boundarycond}; in the following we
use $\lambda = 0.01\Lambda$. Figure \ref{fig:eci_rectwell_wavefunc}
shows the radial wave function of the lowest $l=0$ state with positive
energy for different potential depths $V_0$. Outside the potential the
structure of the wave functions is very similar for the different
interaction strengths. Only the wavelength is changed slightly due to
the different matching to the wave function in the interior (see
insert of Figure \ref{fig:eci_rectwell_wavefunc}). This change of the
relative momentum translates immediately into an energy shift. From
that picture the connection between energy shifts and phase shifts is
evident.

A second point becomes clear from this simple example: The detailed
structure of the radial dependence of the potential or the number of
bound states is irrelevant for the energy shift, only the phase shifts
$\tilde{\eta}_l(q)$ matter. The insert in Figure
\ref{fig:eci_rectwell_wavefunc} shows that the wave functions behave
very different within the range $\lambda$ due to the different
potential depths. Moreover the potentials have a different number of
bound states, e.g., the thick solid curve is associated with a
potential with one bound state but zero phase shift. Hence the
behavior outside the potential is identical to the noninteracting
case (thin solid curve) and the energy shift is zero.

\begin{figure}
\includegraphics[height=0.26\textheight]{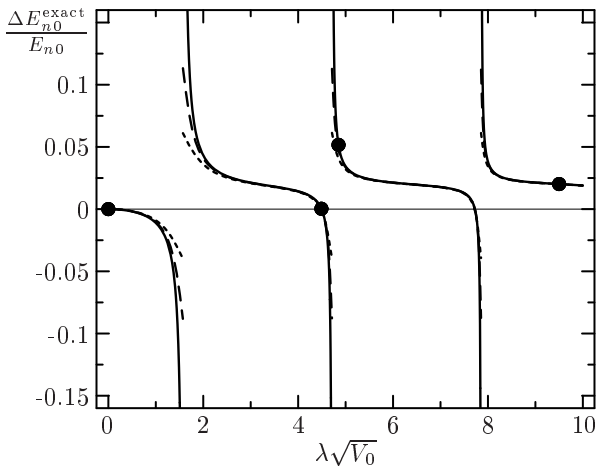}
\caption{Relative energy-shift $\Delta E_{n0}^{\exact}/E_{n0}$ obtained 
  from the exact $l=0$ solutions plotted versus the strength
  $\lambda\sqrt{V_0}$ of the square-well potential. The solid line
  gives the energy shift for the lowest ($n=1$), the dashed line for
  the 10th, and the dotted line for 20th state of the positive energy
  spectrum. The dots mark the interaction strengths used in Figure
  \ref{fig:eci_rectwell_wavefunc}.}
\label{fig:eci_rectwell_energyshift}
\end{figure}

Next we investigate the dependence of the $l=0$ energy shifts on the
strength of the attractive potential. Figure
\ref{fig:eci_rectwell_energyshift} shows the relative energy shift
$\Delta E^{\exact}/E$ versus $\lambda\sqrt{V_0}$, where the radius
$\lambda=0.01\Lambda$ of the square-well potential is kept fixed and
the depth $V_0$ is increased. A characteristic pattern appears: In the
vicinity of interaction strengths where the potential gains another
bound state the relative energy shift assumes large positive and
negative values. Large positive energy shifts occur for potentials
that have a very weakly bound state, negative energy shifts for those
that have an almost bound state. In between these interaction
strengths extended plateaus of nearly constant energy shift
appear. Within the plateaus the energy shift is independent of the
radial quantum number $n$ of the level or the relative momentum. This
is a special property of the s-wave channel; for higher partial waves
the relative energy shift \eqref{eq:eci_energyshift_scattapprox} is
proportional to $q^{2l}$. Only at the edges of the plateaus a slight
dependence on the relative momentum shows up (see Figure
\ref{fig:eci_rectwell_energyshift}).

This structure is closely related to the behavior of the s-wave
scattering length. The $l=0$ energy shift induced by the ECI in
scattering length approximation \eqref{eq:eci_energyshift_scattapprox}
is proportional to the scattering length $a_0$. For the square-well
potential we get
\eq{
  \frac{\Delta E_{n0}}{E_{n0}} = 2 \frac{a_0}{\Lambda}
  \quad\text{with}\quad
  \frac{a_0}{\lambda} = 1 -
  \frac{\tan(\lambda\sqrt{V_0})}{\lambda\sqrt{V_0}} .
}
This ECI energy shift is right on top of the solid curve in Figure
\ref{fig:eci_rectwell_energyshift}, i.e., it agrees very well with the
exact energy shift for the lowest positive energy state. Even for
higher momenta the agreement is very good provided that the magnitude
of the scattering length is not too large. Significant deviations
occur only if momentum \emph{and} scattering length are large.

\begin{figure}
\hskip-2em\includegraphics[height=0.26\textheight]{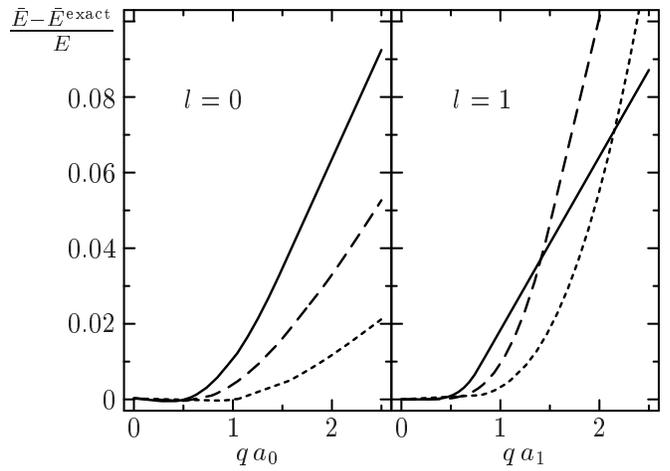}
\caption{Relative deviation of the two-body energy calculated with the
  ECI in scattering length approximation from the exact energy as
  function of $q a_l$ for $l=0$ (left) and $l=1$ states (right). The
  curves were obtained for interactions with three bound states by
  varying the strength $\lambda\sqrt{V_0}$ and looking at the energy
  shifts for different relative momenta $q_{nl}\Lambda\approx20$ (solid),
  $40$ (dashed), and $80$ (dotted).  }
\label{fig:eci_energydeviation}
\end{figure}

To obtain a quantitative measure of for the applicability of the ECI
in scattering length approximation we investigate the relative
deviation $(\bar{E}-\bar{E}^{\exact})/E$ of the ECI energy levels
compared to the exact ones for the square-well potential. We expect
that the agreement gets worse if either the relative momentum or if
the scattering length is large. Therefore, Figure
\ref{fig:eci_energydeviation} shows the relative energy deviation for
$l=0$ and $l=1$ states as function of the product of momentum and
scattering length, $q\,a_l$, which was assumed to be small in order to
introduce the scattering length \eqref{eq:eci_scatteringlength}.
The different curves correspond to different values of the momentum
$q_{nl}$ and were obtained by varying the depth of the square-well potential
and thus the scattering length. 

As expected the deviation increases with increasing value of
$q\,a_l$. Nevertheless, the deviation of the energy calculated with
the ECI in scattering length approximation is below 1\% up to rather
large values of $q\,a_l\lesssim 1$. If we tolerate a maximum deviation
of 5\% then the scattering length formulation can be used up to values
$q\,a_l \lesssim 1.5$.

It should be noted that the relative deviation of the general form
\eqref{eq:eci_energyshift} of the ECI in the parameter range discussed
above is below $10^{-4}$. Thus all approximations made to obtain
equation \eqref{eq:eci_energyshift} are valid on a high level of
accuracy. The restrictions on the validity of the scattering length
formulation \eqref{eq:eci_energyshift_scattapprox} originate for the
replacement of the phase shifts by the scattering length alone, which
is no inherent part of th ECI concept. If the simple scattering length
formulation is not sufficient for a special application one can go
beyond that.

For example, the inclusion of effective volume corrections (see
eq. \eqref{eq:eci_effrange_expansion}) improves the agreement
with the exact energy shifts. In this way we can reduce the maximum
deviation to only 1\% up to $q\,a_0\lesssim1.5$.

\subsection{ECI versus Pseudopotential}

The idea to simulate the effect of a complicated finite-range two-body
potential by a simple s-wave contact interaction dates back to E. Fermi
\cite{Ferm36} and was used by several authors \cite{BlWe79} in various
physical contexts.  K. Huang and C.N. Yang \cite{HuYa57,Huan63}
generalized this idea and constructed the so called
\emph{pseudopotential} that acts in all partial waves.

The aim of the pseudopotential is (a) to generate the phase shifts of
the original potential by a boundary condition at $r=0$ and (b)
to reformulate this by an additional inhomogenous term in the
Schr\"odinger equation of the two-body scattering problem. This
additional term is interpreted as the pseudopotential, which can be
phrased in the following operator form \cite{Roth00}
\eq{ \label{eq:eci_pseudopot_operator}
  \vO^{\pseudo}_l 
  = \int\!\!\dd^3r \ket{\rV}\frac{1}{r^l}\; g^{\pseudo}_l\,\delta^{(3)}(\rV)\; 
    \frac{\overset{\rightarrowtail}{\partial}{}^{2l+1}} 
    {\partial r^{2l+1}} r^{l+1} \bra{\rV}
}
with an interaction strength 
\eq{ \label{eq:eci_pseudopot_strength}
  g^{\pseudo}_l 
  =  \frac{4\pi}{2\mu}\; \frac{(l+1)}{(2l+1)!}\; a_l^{2l+1}  
}
for the $l$-th partial wave. For this discussion we restrict ourselves
to the scattering length approximation of the phase shifts.  Due to
the fact that the radial derivative acts only to the right hand side
the operator of the pseudopotential \eqref{eq:eci_pseudopot_operator}
is not hermitian.  This is in contradiction to the basic concept of
effective interactions.

A more severe weakness shows up when we evaluate the energy shifts
induced by the pseudopotential. As discussed in section
\ref{sec:eci_concept} the expectation value of a proper effective
interaction with eigenstates $\ket{nlm}$ of the noninteracting
two-body system should be equal to the energy shift induced by the
original potential. For the pseudopotential
\eqref{eq:eci_pseudopot_operator} we obtain an energy shift
\eqmulti{ \label{eq:eci_pseudopot_energyshift}
  \frac{\Delta E^{\pseudo}_{nl}}{E_{nl}} 
  &= \frac{1}{E_{nl}}\matrixe{nlm}{\vO^{\pseudo}_l}{nlm} \\
  &= \frac{2}{\Lambda}\; \frac{(l+1)}{[(2l+1)!!]^2}\; 
    q_{nl}^{2l} a_l^{2l+1} .
}
This has to be compared with the full energy shift in scattering
length approximation \eqref{eq:eci_energyshift_scattapprox} which is
by construction reproduced by the ECI. Obviously the energy shift
induced by the pseudopotential for states with $l>0$ is by a factor
$(l+1)/(2l+1)$ smaller the the energy shift of the original
potential. Thus the pseudopotential underestimates the effect of the
two-body interactions beyond s-wave when used in a mean-field
framework. For the widely used s-wave part the energy shifts of the
pseudopotential agree with the energy shifts of the original
potential.

We conclude that the nonhermitian pseudopotential is not a proper
effective interaction for a mean-field description of a dilute
quantum gases that goes beyond s-wave interactions.

\section{Energy Functional of a Trapped Multi-Component Fermi Gas}
\label{sec:ef}

\subsection{Fundamentals}

In the following we investigate the ground state properties of a
dilute Fermi gas composed of $\Xi$ distinguishable components that is
trapped in an external potential $U(\xV)$ at temperature $T=0\,$K. In
present experiments \cite{DeJi99} one or two components are used,
which belong to the same atomic species but are distinguished by
different projections $M_{F}$ of the total angular momentum $F$ onto
the direction of an external magnetic field. We distinguish the
different components by a formal quantum number $\xi=1,\dots,\Xi$. For
simplicity we use the same mass $m$ of the atoms for all components.

We treat the many-body problem in the framework of density functional
theory and construct an energy functional of the inhomogenous
multi-component Fermi gas within a proper approximation. The
ground state density-distribution of the many-body system is then
determined by functional minimization of the energy.

The large particle numbers of the order $N\sim10^6$ allow the rather
simple Thomas-Fermi approximation for the energy functional. It is
assumed that the energy density of the inhomogenous system is
described locally by the energy density of the corresponding
homogenous system, and higher-order terms which include gradients of
the density are small. To check the quality of the Thomas-Fermi
approximation we calculated the next order gradient corrections for a
trapped noninteracting Fermi gas. For $N=100$ particles the relative
contribution of the gradient correction to the total energy is of the
order of $10^{-2}$; for typical particle numbers of $N=10^{6}$ it
drops to $10^{-5}$ \cite{Roth00}.

As starting point for the Thomas-Fermi approximation we calculate the
energy density of the homogenous interacting multi-component Fermi
gas in mean-field approximation. The basic restriction of the
mean-field picture is that two- and many-body correlations induced by
the interaction are not contained in the many-body state. Nevertheless
they can be implemented implicitly by using a proper effective
interaction that is tailored for the model-space available. In the
previous section we constructed the Effective Contact Interaction
especially for the mean-field description of dilute not self-bound
quantum gases. 

A central topic of the following studies is the role of the
interaction on the structure and stability of trapped degenerate Fermi
gases. Our special interest concerns the p-wave part of the
interaction, which contributes even at $T=0$K --- in contrast to
bosonic systems. It will turn out that the p-wave terms can be of
substantial importance for the ground state properties of fermionic
systems and should not be neglected from the outset.

We write the Hamilton operator of the system as a sum of the external
trapping potential $\UO$ and an internal part $\HO_{\text{int}}$
\eq{ \label{eq:ef_hamiltonian}
  \HO = \UO + \HO_{\text{int}}.
}
The internal part contains the kinetic energy and the Effective
Contact Interaction as discussed in section \ref{sec:eci}.  We include
the s-wave and p-wave terms of the ECI in scattering length
formulation as well as the s-wave effective range correction. With
equations \eqref{eq:eci_interactionoperator},
\eqref{eq:eci_interactionstrength_scattapprox},
\eqref{eq:eci_interactionoperator_expansion}, and
\eqref{eq:eci_interactionstrength_effvolume} the internal Hamiltonian
reads
\eqalignat{2}{ \label{eq:ef_hamiltonian_int}
  \HO_{\text{int}}
  &= \;\frac{1}{2m} &&\: \sum_i \pOV_i^2 \nonumber \\ 
  &+ \;\frac{4\pi}{m}\,a_0 && \sum_{i,\,j>i} \delta^{(3)}(\rOV_{ij}) \\ 
  &- \frac{12\pi}{m}\,b_0  && \sum_{i,\,j>i} \frac{1}{2}\big[
    \delta^{(3)}(\rOV_{ij}) (\rUOV_{ij}\!\cdot\!\qOV_{ij})^2 
    + \text{h.a.}\big] \nonumber \\
  &+ \frac{12\pi}{m}\,a_1^3 && \sum_{i,\,j>i}  
    (\qOV_{ij}\!\cdot\!\rUOV_{ij})\;\delta^{(3)}(\rOV_{ij})\;
    (\rUOV_{ij}\!\cdot\!\qOV_{ij}) . \nonumber
}
The summations over the particle indices $i$ and $j$ range from $1$ to
the total number of particles. The properties of the two-body
interaction are parameterized by the s- and p-wave scattering lengths
$a_0$ and $a_1$, respectively, and by the s-wave effective volume
$b_0$. In general the interaction parameters depend on the component
quantum numbers $\xi$ of the interacting particles. In order to
discuss the basic phenomena we restrict ourselves to equal interaction
parameters for all components.  The generalization to scattering
length matrices that account for the dependence on the component indices
of the two interacting particles is straightforward.

Experimentally each component may experience a different trapping
potential $U_{\xi}(\xV)$. For magnetic traps this is due to the
different magnetic momenta of the components, which leads to a
relative shift of the trapping potentials for the components. Thus the
operator of the external potential has the following form
\eq{
  \UO = \sum_i \sum_{\xi} U_{\xi}(\xOV_i)\,\PiO_{\xi,i}
}
where $\PiO_{\xi}$ is a projection operator onto states with the
component quantum number $\xi$.

Many of the results shown in the next sections do not depend on the
actual shape of the trapping potential. If the shape enters explicitly
we assume a deformed harmonic oscillator potential
\eqmulti{ \label{eq:ef_trappingpot}
  U(\xV) 
  &= \frac{m\omega^2}{2} \big(\lambda_1^2 x_1^2 
    + \lambda_2^2 x_2^2 + \lambda_3^2 x_3^2 \big) \\ 
  &= \frac{1}{2m\ell^4} \big(\lambda_1^2 x_1^2 
    + \lambda_2^2 x_2^2 + \lambda_3^2 x_3^2 \big) ,
}
where $\omega = \sqrt[3]{\omega_1\omega_2\omega_3}$ is the mean
oscillator frequency and $\ell = (m\omega)^{-1/2}$ the corresponding
mean oscillator length, i.e., the mean width of the Gaussian
single-particle ground state of the harmonic oscillator potential. The
deformation is parameterized by the ratios $\lambda_i =
\omega_i/\omega$, which fulfill the condition $\lambda_1 \lambda_2
\lambda_3 = 1$.

\subsection{Energy Density in Thomas-Fermi Approximation}

The calculation of the energy density functional of the inhomogenous
interacting Fermi gas is performed in two steps: First we calculate
the energy density of the corresponding homogenous system in
mean-field approximation. In the second step this is translated into
an energy density of the inhomogenous system by means of the
Thomas-Fermi approximation.

The ground state of a many-fermion system in mean-field approximation
is given by an antisymmetrized product of one-body states
$\ket{i}$. In the case of a homogenous system the one-body states
are eigenstate of the momentum operator with eigenvalues $\kV_i$. In
addition they are characterized by the component quantum number $\xi$
\eq{
  \ket{i} = \ket{\kV_i} \otimes \ket{\xi_i} .
}
Assuming a box of volume $V$ with periodic boundary conditions the
spatial part of the one-body states is given by  
\eq{
  \braket{\xV}{\kV_i} 
  = \frac{1}{\sqrt{V}} \exp\!\big(\ii\, \kV_i\! \cdot\! \xV\big) .
}

The energy density of the homogenous system is given by the
expectation value of the internal part of the Hamilton operator
\eqref{eq:ef_hamiltonian_int}
\eq{
  \EC_{\text{hom}} = \expect{\HO_{\text{int}}}/V .
}
The calculation of the expectation values of the several parts of the
Hamiltonian is straightforward \cite{Roth00}. As a function of the
Fermi momenta $\kappa_{\xi}$ of the different components the
energy density reads
\eqalignat{2}{ \label{eq:ef_energydens_hom}
  &\EC_{\text{hom}}(\kappa_1,\dots,\kappa_{\Xi} &&) \nonumber\\
  &\qquad= \frac{1}{20 \pi^2 m} &&\;\: \sum_{\xi} \;\:
    \kappa_{\xi}^5 \nonumber \\
  &\qquad+ \;\frac{a_0}{9 \pi^3 m} && \sum_{\xi,\,\xi'>\xi} \kappa_{\xi}^3
    \kappa_{\xi'}^3 \\
  &\qquad+ \frac{a_1^3}{30 \pi^3 m} && \;\: \sum_{\xi} \;\:
    \kappa_{\xi}^8   \nonumber \\
  &\qquad+ \frac{a_1^3+b_0}{60 \pi^3 m} && \sum_{\xi,\,\xi'>\xi}
    [ \kappa_{\xi}^3 \kappa_{\xi'}^5 
    + \kappa_{\xi}^5 \kappa_{\xi'}^3 ]  \nonumber . 
}
The summations run over all components $\xi=1,\dots,\Xi$. To avoid
fractional exponents we use Fermi momenta $\kappa_{\xi}$ rather than
densities $\rho_{\xi}=\kappa_{\xi}^3/(6\pi^2)$.

The basic assumption of the Thomas-Fermi (or local-density)
approximation is that the energy density of the inhomogenous Fermi
gas is locally given by the energy density of the corresponding
homogenous system. Thus the energy density of the inhomogenous
system is constructed from \eqref{eq:ef_energydens_hom} by replacing
$\kappa_{\xi}$ with \emph{local Fermi momenta} $\kappa_{\xi}(\xV)$. In
addition the contribution of the external trapping potential has to be
included. This results in the following expression for the
energy density of the trapped interacting multi-component Fermi gas
\eqalignat{2}{ \label{eq:ef_energydens}
  &\EC[\kappa_1,\dots,\kappa_{\Xi}](\xV) && \nonumber \\
  &\qquad= \;\;\;\frac{1}{6\pi^2} && \;\: \sum_{\xi} \;\:
    U_{\xi}(\xV)\; \kappa_{\xi}^3(\xV) \nonumber \\
  &\qquad+ \;\frac{1}{20 \pi^2 m} && \;\: \sum_{\xi} \;\:
    \kappa_{\xi}^5(\xV) \nonumber \\
  &\qquad+ \;\frac{a_0}{9 \pi^3 m} && \sum_{\xi,\,\xi'>\xi}
    \kappa_{\xi}^3(\xV) \kappa_{\xi'}^3(\xV) \\  
  &\qquad+ \frac{a_1^3}{30 \pi^3 m} && \;\: \sum_{\xi} \;\:
    \kappa_{\xi}^8(\xV) \nonumber \\
  &\qquad+ \frac{a_1^3+b_0}{60 \pi^3 m} && \sum_{\xi,\,\xi'>\xi} 
    [ \kappa_{\xi}^3(\xV) \kappa_{\xi'}^5(\xV) 
    + \kappa_{\xi}^5(\xV) \kappa_{\xi'}^3(\xV) ]  . \nonumber
}
The local Fermi momentum is related to the density of particles of the
component $\xi$ by
\eq{
  \rho_{\xi}(\xV) = \frac{1}{6\pi^2}\,\kappa_{\xi}^3(\xV) .
}
Accordingly the number of particles of component $\xi$ is given by
\eq{ \label{eq:ef_particlenumber}
  N_{\xi} 
  =  \int\!\!\dd^3x\;\rho_{\xi}(\xV)
  = \frac{1}{6\pi^2} \int\!\!\dd^3x\;\kappa_{\xi}^3(\xV) .
}

As discussed in section \ref{sec:eci_example} we can reproduce the
two-body energy spectrum with an accuracy of about 5\% up to $a_l q
\approx 1.5$. If we take this as a limit for the root mean square of
the relative momentum $\langle\qO^2\rangle^{1/2} = 0.53\,\kappa$ in the
many-body system, we can apply our many-body model up to $a_l\kappa
\approx 3$.

\subsection{Functional Variation and the Extremum Condition}
\label{sec:ef_funcvar}

The ground state density of a system is found by minimizing the energy
functional
\eq{
  E[\kappa_1,\dots,\kappa_{\Xi}] 
  = \int\!\!\dd^3x\; \EC[\kappa_1,\dots,\kappa_{\Xi}](\xV) .
}
for given particle numbers $N_{\xi}$. This constraint is implemented
with help of a set of Lagrange multipliers $\mu_{\xi}$, which are the
chemical potentials of the different components. The Legendre
transformed functional
\eqmulti{ \label{eq:ef_energydens_f}
  &F[\kappa_1,\dots,\kappa_{\Xi}] \\
  &\qquad= E[\kappa_1,\dots,\kappa_{\Xi}] - \sum_{\xi} \mu_{\xi} N_{\xi} \\[-3pt]
  &\qquad= \int\!\!\dd^3x\; \EC[\kappa_1,\dots,\kappa_{\Xi}](\xV) 
   - \sum_{\xi}\frac{\mu_{\xi}}{6\pi^2}\kappa_{\xi}^3(\xV) \\
  &\qquad= \int\!\!\dd^3x\; \FC[\kappa_1,\dots,\kappa_{\Xi}](\xV) ,
}
has to be minimized by functional variation. A necessary but not
sufficient condition for a set of local Fermi momenta
$\{\kappa_1(\xV),\dots,\kappa_{\Xi}(\xV)\}$ to minimize the
transformed energy functional $F[\kappa_1,\dots,\kappa_{\Xi}]$ is
stationarity, i.e., that the first variation of
$F[\kappa_1,\dots,\kappa_{\Xi}]$ with respect to all
$\kappa_{\xi}(\xV)$ vanishes
\eq{
  \frac{\delta}{\delta\kappa_{\xi}} F[\kappa_1,\dots,\kappa_{\xi}] = 0
  \qquad\text{for all}\quad \xi.
}
This extremum condition is fulfilled if the derivative of the integrand
$\FC[\kappa_1,\dots,\kappa_{\Xi}](\xV)$ with respect to all local
Fermi momenta vanishes at each point $\xV$
\eq{
  \frac{\partial}{\partial\kappa_{\xi}(\xV)} 
  \FC[\kappa_1,\dots,\kappa_{\Xi}](\xV) = 0
  \qquad\text{for all}\quad\xV,\,\xi .
}
Inserting expression \eqref{eq:ef_energydens} for the energy density
and evaluating the derivative results in the extremum condition
\eqmulti{ \label{eq:ef_extremumcond}
  &m[\mu_{\xi}-U_{\xi}(\xV)]  \\
  &\qquad = \frac{1}{2}\kappa_{\xi}^2(\xV) 
    + \frac{2 a_0}{3\pi} \sum_{\xi'\ne\xi} \kappa_{\xi'}^3(\xV)  
    + \frac{8 a_1^3}{15\pi} \kappa_{\xi}^5(\xV)  \\
  &\qquad  + \frac{a_1^3+b_0}{30\pi} \sum_{\xi'\ne\xi} [
    3\,\kappa_{\xi'}^5(\xV) + 5\,\kappa_{\xi}^2(\xV)\,\kappa_{\xi'}^3(\xV) ]
}
for all $\xV$ and each $\xi$. This is a coupled set of $\Xi$
polynomial equations for the local Fermi momenta
$\{\kappa_1(\xV),\dots,\kappa_{\Xi}(\xV)\}$ at some given point
$\xV$. Note that the trivial solutions $\kappa_{\xi}(\xV)=0$ were
separated already. Any real solution of the extremum condition
\eqref{eq:ef_extremumcond} corresponds to a stationary point of the
energy functional. In general one has to check explicitly whether they
correspond to a minimum of the energy functional or whether they are
maxima or saddle points.

All following investigations on the structure and stability of
degenerate Fermi gases and on the influence of s- and p-wave
interactions are based on the extremum condition
\eqref{eq:ef_extremumcond}. Many physical conclusions can be drawn
from its algebraic structure already. We will discuss these questions
in detail for the one- and two-component Fermi gas in section
\ref{sec:1comp} and \ref{sec:2comp}, respectively.

\section{Single-Component Fermi Gas}
\label{sec:1comp}

As a first application of the formalism developed in the preceding
section we study the properties of a degenerate single-component
Fermi gas. 

\subsection{Effect of the p-Wave Interaction}

The energy density of the interacting multi-component Fermi gas
\eqref{eq:ef_energydens} reduces for the single-component system to
the form
\eqmulti{ \label{eq:1comp_energydens}
  \EC[\kappa](\xV) 
  &= \frac{1}{6\pi^2}\,U(\xV)\, \kappa^3(\xV)
    + \frac{1}{20\pi^2m}\,\kappa^5(\xV) \\
  &+ \frac{a_1^3}{30\pi^3m}\,\kappa^8(\xV) ,
}
where $\kappa(\xV)$ is the local Fermi momentum. The first term is the
contribution of the trapping potential $U(\xV)$, the second term is
the kinetic energy, and the third term describes the contribution of
the p-wave interaction with a p-wave scattering length $a_1$. As
mentioned earlier the s-wave part of the interaction does not
contribute in a system of identical fermions due to the Pauli
principle. Therefore the p-wave part is the leading interaction term
and there is no reason to neglect it from the outset. 
 
A first hint on the effects of the p-wave interaction are given by the
density distributions for different values of the p-wave scattering
length. The density distribution is obtained by solution of the
extremum condition \eqref{eq:ef_extremumcond}, which takes the simple
form
\eqgather{ \label{eq:1comp_extremumcond}
  m[\mu-U(\xV)] = f_1[\kappa(\xV)] \\[4pt]
  \text{with}\quad
  f_1(\kappa) = \frac{1}{2}\,\kappa^2 
    + \frac{8 a_1^3}{15\pi}\,\kappa^5 \nonumber .
}
This 5th order polynomial equation for the local Fermi momentum
$\kappa(\xV)$ is solved numerically for each point $\xV$. 
The chemical potential $\mu$ is adjusted such that particle number
\eqref{eq:ef_particlenumber} assumes the desired value.

\begin{figure}
\includegraphics[height=0.27\textheight]{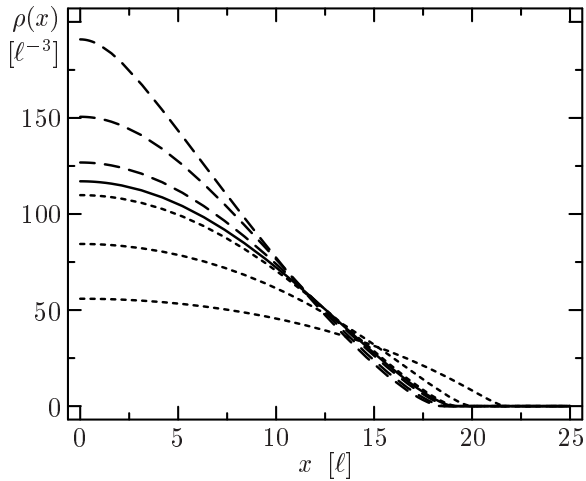}
\caption{Density profile $\rho(x)$ of a single-component 
  Fermi gas of $N=10^6$ particles trapped in a spherical symmetric parabolic
  trap with oscillator length $\ell$. The solid curve shows the
  noninteracting gas $a_1/\ell=0$. Dotted curves correspond to
  repulsive p-wave interactions with $a_1/\ell=0.03$,
  $0.06$, and $0.1$ (from top to bottom). The dashed curves 
  show attractive p-wave interactions with
  $a_1/\ell=-0.03$, $-0.04$ and $-0.044$ (top to bottom), respectively. }
\label{fig:1comp_density}
\end{figure}

Figure \ref{fig:1comp_density} shows the resulting radial density
profiles $\rho(x) = \kappa^3(x)/(6\pi^2)$ for a single-component gas
of $N=10^6$ particles in a spherical trap with oscillator length
$\ell$ for different p-wave scattering lengths $a_1$. The oscillator
length defines the fundamental length scale of the problem and the
parameter that determines the strength of the interaction is the ratio
of the p-wave scattering length and oscillator length, $a_1/\ell$. To
increase the magnitude of this ratio experimentally one can either
increase the magnitude of the scattering length or decrease the
oscillator length.

For a repulsive p-wave interaction, i.e., $a_1/\ell>0$, of
increasing strength (dotted curves) the density distribution flattens
and expands radially compared to the noninteracting system (solid
line). For a ratio $a_1/\ell=0.1$ the central density has dropped to
one half of the density of the noninteracting gas. With a typical
experimental oscillator length of $\ell=1\mu m$ this ratio corresponds
to a rather large scattering length of $a_1 \approx 2000
a_{\text{B}}$, which nevertheless may be within the range of
experimental parameters \cite{Bohn00}. For a tightly confining trap
with $\ell=0.1\mu m$ a moderate scattering length of $a_1\approx 200
a_{\text{B}}$ is required to obtain the same ratio.

For an attractive p-wave interaction, $a_1/\ell<0$, the central
density increases significantly with increasing interaction
strength. If the central density exceeds a certain value, i.e., if
$|a_1/\ell|$ exceeds a critical value, then the extremum condition
\eqref{eq:1comp_extremumcond} has no real solution any more.
Physically this corresponds to a collapse of the dilute gas caused by
the attractive mean-field that is generated by the p-wave
interaction. We will discuss this question in detail in the following
sections.

The dependence of the density distribution on the p-wave scattering
length as depicted in Figure \ref {fig:1comp_density} already
demonstrates that the p-wave interaction may have strong influence on
the properties of degenerate Fermi gases.

\subsection{Mean-Field Instability: A Variational Picture}

To illustrate the origin and mechanism of the collapse of the
metastable state of the trapped Fermi gas we utilize a simple
variational picture. Assume a single-component Fermi gas of $N$
particles in a spherical symmetric oscillator potential.  The local
Fermi momentum of the interacting system is parameterized by the
analytic expression for the local Fermi momentum of the noninteracting
system
\eq{
  \kappa(\xV) 
  = \frac{2(6N)^{1/3}}{X_{\text{t}}} 
  \sqrt{1-\Big(\frac{x}{X_{\text{t}}}\Big)^2}
  \qquad\text{for}\;x\le X_{\text{t}} ,
}
where the classical turning point $X_{\text{t}}$ is treated as
variational parameter. By inserting this parameterization into the
energy density \eqref{eq:1comp_energydens} and integrating we obtain
a closed expression for the energy as function of the parameter $X_{\text{t}}$
\eq{ \label{eq:1comp_var_energy}
  E(X_{\text{t}}) 
  = C_u \frac{N X_{\text{t}}^2}{\ell^4}
  + C_t \frac{N^{5/3}}{X_{\text{t}}^2}
  + C_1 \frac{N^{8/3}\,a_1^3}{X_{\text{t}}^5} ,
}
with constant coefficients 
\eq{
  C_u = \frac{3}{16m}\,,\;\;
  C_t = \frac{3(9/2)^{1/3}}{2m}\,,\;\;
  C_1 = \frac{8^6 (4/3)^{1/3}}{1925\pi^2m} . 
}
Again the first term corresponds to the external potentials, the
second to the kinetic energy, and the third term to the p-wave
interaction.

\begin{figure}
\includegraphics[height=0.27\textheight]{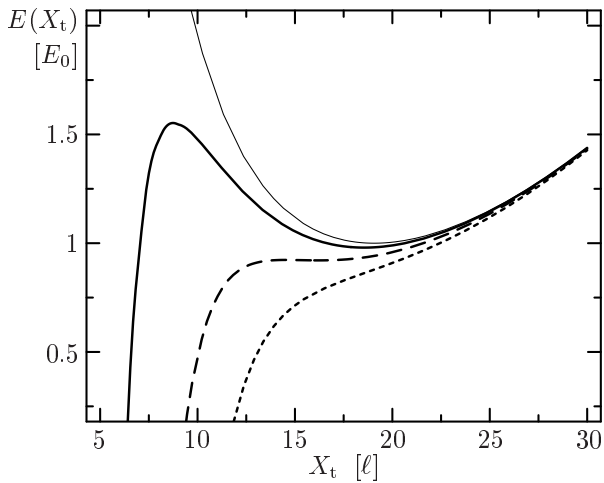}
\caption{Variational energy \eqref{eq:1comp_var_energy} of a trapped 
  single-component Fermi gas with $N=10^6$ as function of the parameter
  $X_{\text{t}}$. The curves show the noninteracting gas (thin
  solid), $a_1/\ell = -0.035$ (solid), $a_1/\ell = -0.051$ (dashed),
  and $a_1/\ell = -0.065$ (dotted). The energies are given in units of
  the ground state energy $E_0$ of the noninteracting gas.}
\label{fig:1comp_var_energy}
\end{figure}

Figure \ref{fig:1comp_var_energy} shows the dependence of the total
energy \eqref{eq:1comp_var_energy} on the parameter $X_{\text{t}}$ for
a system of $N=10^6$ particles with different p-wave scattering
lengths.  For attractive p-wave interactions, i.e., negative
scattering length $a_1$, the contribution of the interaction in
\eqref{eq:1comp_var_energy} is negative.  Due to its
$X_{\text{t}}^{-5}$ dependence this interaction contribution leads to
a rapid drop of the energy for systems of decreasing spatial extension
and thus increasing density. At very high densities (small
$X_{\text{t}}$) one formally ends up with states of negative energy,
i.e., bound states. One should however keep in mind that the
assumptions made for the construction of the Effective Contact
Interaction are not valid in this high density regime any more.

The not self-bound metastable state appears as local minimum at
positive energies and low densities provided the p-wave attraction is
sufficiently weak; the thick solid curve in Figure
\ref{fig:1comp_var_energy} shows an example. If the strength of the
attractive p-wave interaction increases then the local minimum
flattens and devolves to a saddle point (dashed curve). From this
particular interaction strength on the metastable low-density state
does not exist anymore, only the true ground state of the system
remains, which is usually a crystal. The system collapses if the
barrier caused by the positive kinetic and the attractive mean-field
energy vanishes. Since the mean-field attraction grows with increasing
density the system is unstable and collapses towards a high-density
configuration.

\subsection{Mean-Field Instability: Stability Conditions}

Based on the extremum condition \eqref{eq:1comp_extremumcond} we
derive a set of analytic stability conditions that relate the maximum
density of a metastable system with the p-wave scattering length. Part
of this was already discussed in \cite{RoFe00b}.

\begin{figure}
\includegraphics[height=0.27\textheight]{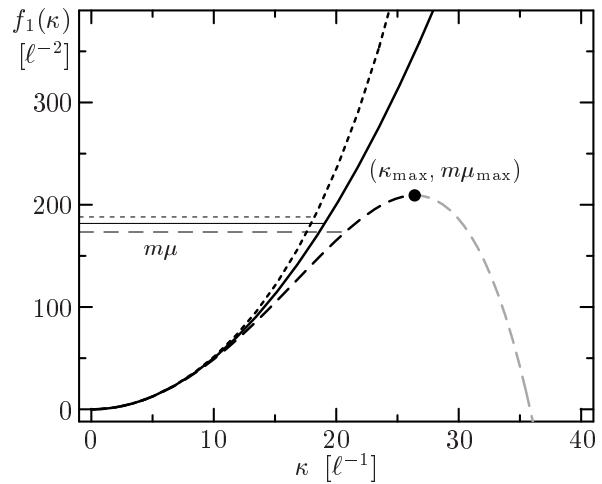}
\caption{Right hand side $f_1(\kappa)$ of the extremum condition
  \eqref{eq:1comp_extremumcond} as function of the Fermi momentum for
  a noninteracting singe-component gas (solid), with repulsive p-wave
  interaction $a_1/\ell=0.04$ (dotted), and with attractive p-wave 
  interaction $a_1/\ell=-0.04$ (dashed). The horizontal lines mark the
  respective values of the chemical potentials for $N=10^6$ particles.}
\label{fig:1comp_extremumcond}
\end{figure}

The mean-field instability of the system occurs at values of the
chemical potential $\mu$ and the scattering length $a_1$, where the
extremum condition \eqref{eq:1comp_extremumcond} does not have a real
solution any more. This is shown in a pictorial way in Figure
\ref{fig:1comp_extremumcond} where the the right hand side
$f_1(\kappa)$ of the extremum condition is plotted as function of
$\kappa$ for different $a_1/\ell$. The solution of the extremum
condition at some specific point $\xV$ is given by the value of
$\kappa$ at which the respective curve reaches the value
$m[\mu-U(\xV)]$.  In the minimum of the trapping potential (we assume
$U(\xV)=0$ in the minimum) the solution is given by the point where
$f_1(\kappa)$ reaches the value $m\mu$. By moving towards the outer
regions of the trap $m[\mu-U(\xV)]$ decreases and one scans
$f_1(\kappa)$ down along ordinate until one reaches $m[\mu-U(\xV)] =
0$, i.e., the classical turning point.

For repulsive p-wave interactions (dotted curve) the r.h.s. of
\eqref{eq:1comp_extremumcond} is a monotonic growing function and 
solutions exist for arbitrary values of $m[\mu-U(\xV)]$. If the
scattering length $a_1$ is negative (dashed curve) then $f_1(\kappa)$
exhibits a maximum at a Fermi momentum $\kappa_{\text{max}}$ and
chemical potential $\mu_{\text{max}}$
\eq{
  \kappa_{\text{max}} = -\frac{\sqrt[3]{3\pi}}{2 a_1}
  \,,\qquad
  \mu_{\text{max}} = \frac{3(3\pi)^{3/2}}{40 m a_1^2} .
}
For values of the chemical potential $\mu>\mu_{\text{max}}$ no
solution of the extremum condition exists, i.e., there is no
metastable low-density state. Equivalently only solutions with local
Fermi momenta below $\kappa_{\text{max}}$ correspond to minima of the
energy functional; those above $\kappa_{\text{max}}$ (gray segment of
the dashed curve) correspond to maxima of the energy. Thus we get a
limiting condition for the local Fermi momentum of the metastable
state
\eq{ \label{eq:1comp_stabilitycond_kF}
  -a_1 \kappa(\xV) \le \frac{\sqrt[3]{3\pi}}{2}
}
or in terms of the density
\eq{ \label{eq:1comp_stabilitycond_rho}
  -a_1^3 \rho(\xV) \le \frac{1}{16\pi} .
}
This is one form of the \emph{stability condition} for the
single-component Fermi gas. We note that this condition is completely
independent of the trap geometry. As soon as the stability condition
is violated somewhere in the trap, in general in the minimum of the
trapping potential, the system will become unstable.

For practical purposes we formulate a stability condition in terms of
the particle number $N$. The maximum particle number $N_{\text{max}}$
of the metastable degenerate Fermi gas is directly connected to the
maximum chemical potential $\mu_{\text{max}}$. This relation is
established numerically by solving the extremum condition for the
maximum chemical potential $\mu_{\text{max}}$ and integrating over the
resulting density distribution to obtain the corresponding maximum
particle number \eqref{eq:ef_particlenumber}. This is done for several
scattering length $a_1$ assuming a deformed oscillator potential
\eqref{eq:ef_trappingpot} with mean oscillator length $\ell$. Finally
a parametrized form of the stability condition is fitted to this data.
The parameterization is motivated by the noninteracting gas, where the
maximum local Fermi momentum is proportional to
$\sqrt[6]{N}/\ell$. Inserting this into the stability condition
\eqref{eq:1comp_stabilitycond_kF} leads to the form
\eq{ \label{eq:1comp_stabilitycond_N}
C\, \Big(\! \sqrt[6]{N}\, \frac{a_1}{\ell}\Big) \le 1
\,,\qquad
C = -2.246 .
}
The parameter $C$ is fitted to the numerical results, which are
reproduced with a deviation far below 1\%. Note that this condition
is independent of the deformation of the harmonic oscillator trap
\cite{Roth00}.

For an interaction strength of $a_1/\ell=-0.01$ which corresponds to a
scattering length $a_1\approx -200a_{\text{B}}$ for a trap with
$\ell=1\mu m$ the maximum particle number is $N_{\text{max}} =
7.8\times10^9$. This seems to be out of the range of present
experiments. Nevertheless if we increase the strength of the p-wave
attraction to $a_1/\ell=-0.1$ then the maximum particle number drops
to $N_{\text{max}} = 7800$. Experimentally this could be achieved by
utilizing a p-wave Feshbach resonance to increase the p-wave
scattering length to $a_1\approx -2000a_{\text{B}}$ as proposed by
J. Bohn \cite{Bohn00} for the ${}^{40}$K system.

\section{Two-Component Fermi Gas}
\label{sec:2comp}

As second application we consider the degenerate two-component Fermi
gas. 

\subsection{Interplay between s- and p-Wave Interaction}

The general energy density of a trapped multi-component Fermi gas in
Thomas-Fermi approximation \eqref{eq:ef_energydens} takes for the
two-component system the following form 
\eqalignat{2}{ \label{eq:2comp_energydens_1}
  \EC[\kappa_1,\kappa_2](\xV)  
  &= \;\;\frac{1}{6\pi^2} && \big[ U_1(\xV) \kappa_1^3(\xV) 
    +  U_2(\xV) \kappa_2^3(\xV) \big] \nonumber \\
  &+ \frac{1}{20\pi^2m} && \big[ \kappa_1^5(\xV) 
    + \kappa_2^5(\xV) \big] \nonumber \\[4pt]
  &+ \;\frac{a_0}{9\pi^3m} && \:\kappa_1^3(\xV) \kappa_2^3(\xV) \\
  &+ \frac{a_1^3}{30\pi^3m} && \big[\kappa_1^8(\xV) 
    + \kappa_2^8(\xV)] \nonumber \\[2pt] 
  &+ \frac{a_1^3+b_0}{60\pi^3m} && \big[ \kappa_1^3(\xV) \kappa_2^5(\xV) 
    + \kappa_1^5(\xV) \kappa_2^3(\xV) \big] \nonumber ,  
}
where $\kappa_1(\xV)$ and $\kappa_2(\xV)$ denote the local Fermi
momenta of the two components. In contrast to the single-component
system, both, s- and p-wave terms of the Effective Contact
Interaction contribute. The s-wave interaction acts only between
particles of different species and generates a contribution
proportional to the product of the densities of both components in the
energy density \eqref{eq:2comp_energydens_1}. The p-wave term acts
between particles of different components as well as between particles
of the same species. For reasons of simplicity we assume the same
p-wave scattering length $a_1$ for these different interactions. 

Including the constraint of given particle numbers $N_1$ and $N_2$ of
the two components with help of the chemical potentials $\mu_1$ and
$\mu_2$ (see section \ref{sec:ef_funcvar}) leads to the transformed
energy density
\eqmulti{ \label{eq:2comp_energydens_f_1}
  \FC[\kappa_1,\kappa_2](\xV) 
  &=  \EC[\kappa_1,\kappa_2](\xV) \\
  &- \frac{\mu_1}{6\pi^2}\kappa_1^3(\xV) 
     - \frac{\mu_2}{6\pi^2}\kappa_2^3(\xV) .
}
Functional variation of the transformed energy functional leads to the
extremum condition. For the two-component system the general form 
\eqref{eq:ef_extremumcond} reduces to a coupled set of two polynomial
equations
\eqmulti{ \label{eq:2comp_extremumcond_1}
  m[\mu_1 - U_1(\xV)] 
  &= \frac{1}{2} \kappa_1^2(\xV)
    + \frac{2 a_0}{3\pi} \kappa_2^3(\xV) 
    + \frac{8 a_1^3}{15\pi} \kappa_1^5(\xV) \\ 
  &+ \frac{a_1^3+b_0}{30\pi} \big[ 3 \kappa_2^5(\xV) 
    + 5 \kappa_1^2(\xV) \kappa_2^3(\xV) \big] ,
}
where the second equation is generated by the exchange
$\kappa_1(\xV)\leftrightarrow\kappa_2(\xV)$ and
$[\mu_1-U_1(\xV)]\rightarrow[\mu_2-U_2(\xV)]$. Trivial solutions
with $\kappa_1(\xV)=0$ and $\kappa_2(\xV)=0$, resp., are already
separated in this expression.

These coupled equations have a great variety of solutions. In order to
show the generic phenomena of the two-component system without too
many parameters we restrict ourselves to equal numbers of particles in
both components $N=N_1=N_2$ as well as trapping potentials that differ
only by an additive constant, thus $\mu-U(\xV)=\mu_1-U_1(\xV)=\mu_2-U_2(\xV)$.

We will concentrate the further studies on the stability of the
degenerate two-component Fermi gas against mean-field collapse. For
this phenomenon solutions with identical local Fermi momenta for both
components, $\kappa(\xV) =\kappa_1(\xV) = \kappa_2(\xV)$, are relevant.  
Under this assumption the extremum condition reduces to a single equation
\eqgather{ \label{eq:2comp_extremumcond_2} 
  m[\mu - U(\xV)] = f_2[\kappa(\xV)] \\[4pt]
\text{with}\quad
  f_2(\kappa) = \frac{1}{2} \kappa^2 
   + \frac{2 a_0}{3\pi} \kappa^3 
   + \frac{4 \tilde{a}_1^3}{5\pi} \kappa^5 \nonumber .
}
For simplicity we introduce a modified p-wave scattering length
\footnote{This can be generalized to include different p-wave
scattering lengths for the different combinations of the two species:
$\tilde{a}_1^3= \frac{1}{3}(a_{1[11]}^3 + a_{1[22]}^3 + a_{1[12]}^3
+b_0 )$.}
\eq{
  \tilde{a}_1^3 = a_1^3 + b_0/3 ,
}
which contains the s-wave effective volume parameter. In the following
we will discuss the properties of the two-component Fermi gas as
function of the s-wave and the modified p-wave scattering length.

For other phenomena, like the separation of the two components due to
repulsive interactions, different classes of solutions become
important. We will discuss these in a forthcoming publication.

\subsection{Mean-Field Instability: Stability Conditions}

The stability of the two-component Fermi gas under the influence of s-
and p-wave interactions can be investigated with tools similar to
the single-component case. Here the solutions with identical Fermi
momenta for both components are of interest.  

Similar to the single-component case the right hand side $f_2(\kappa)$
of the extremum condition \eqref{eq:2comp_extremumcond_2} may exhibit
a maximum if the s-wave or the p-wave scattering length is
negative. Thus the density and particle number of the metastable
low-density state may be limited. For a detailed analysis one has to
look at all possible combinations of signs of the s- and p-wave
scattering lengths separately:
\begin{description}
\item[$a_0\ge0, \tilde{a}_1\ge0$]: For a purely repulsive interaction 
  $f_2(\kappa)$ is a monotonic growing function and no mean-field 
  induced collapse occurs.
\item[$a_0<0, \tilde{a}_1\le0$]: For purely attractive interactions
  $f_2(\kappa)$ shows a maximum; thus the density of the metastable
  low-density state is limited.
\item[$a_0\ge0, \tilde{a}_1<0$]: The negative contribution of the p-wave
  interaction dominates $f_2(\kappa)$ at high densities and generates a
  maximum, i.e., the mean-field induced collapse can occur even if the
  s-wave interaction is repulsive.
\item[$a_0<0, \tilde{a}_1>0$]: It depends on the relative strength of the s-
  and p-wave interaction whether the r.h.s. of the extremum condition
  has a local maximum or grows monotonically.
\end{description}
Especially the stability in the last case depends on a subtle
competition between s- and p-wave interaction. Moreover it shows some
completely new phenomena, which will be discussed in the following
section.
  
For those cases where $f_2(\kappa)$ has a maximum the value of the local
Fermi momentum $\kappa_{\text{max}}$ at the maximum
is given by the equation
\eq{ \label{eq:2comp_stabilitycond_kFmax}
  -a_0 \kappa_{\text{max}} - 2 [\tilde{a}_1 \kappa_{\text{max}}]^3 
  = \frac{\pi}{2} .
}
Again $\kappa_{\text{max}}$ is an upper limit for the local Fermi
momenta, which can occur for a metastable low-density state of the
two-component gas. Thus we can formulate the stability condition
\eq{ \label{eq:2comp_stabilitycond_kF}
  -a_0 \kappa(\xV) - 2 [\tilde{a}_1 \kappa(\xV)]^3 \le \frac{\pi}{2}
}
or equivalently in terms of the density
\eq{ 
  -[6\pi^2\,a_0^3\rho(\xV)]^{1/3} - 12 \pi^2\, \tilde{a}_1^3\rho(\xV) \le
  \frac{\pi}{2} .
}
If these stability conditions are violated than no metastable
low-density state exists for the two-component Fermi gas.  For a pure
s-wave interaction ($\tilde{a}_1=0$) the stability condition
\eqref{eq:2comp_stabilitycond_kF} reduces to the form $-a_0
\kappa(\xV)\le \pi/2$, which was obtained earlier by M. Houbiers
\emph{et. al.} \cite{HoFe97}. Compared to this simple form the
inclusion of the p-wave interaction reveals several new effects.

\begin{figure}
\includegraphics[height=0.27\textheight]{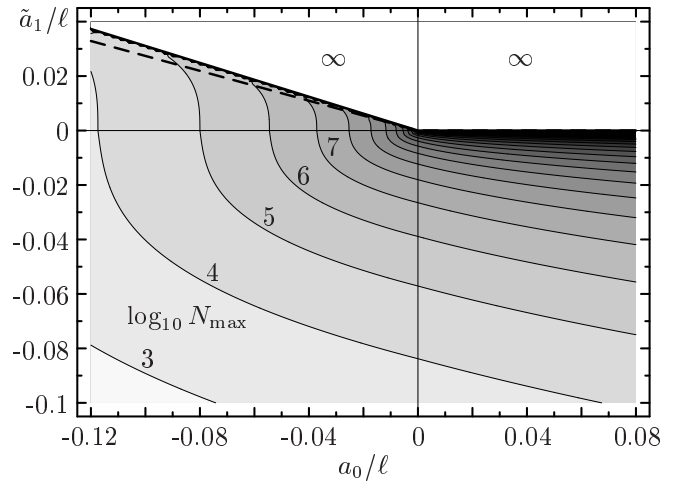}
\caption{Contour plot of the logarithm of the maximum particle number,
  $\log_{10}N_{\text{max}}$, as function of the s- and p-wave
  scattering lengths for a two-component Fermi gas in a harmonic
  oscillator potential with mean oscillator length $\ell$. Selected
  contours are labeled with the corresponding value of
  $\log_{10}N_{\text{max}}$.  In the white area at positive p-wave
  scattering length no collapse can occur, i.e., the maximum particle
  number is infinity.}
\label{fig:2comp_stabmap}
\end{figure}

\begin{table}
\caption{Parameters of the fitted stability condition 
  \eqref{eq:2comp_stabilitycond_N} for the two-component Fermi gas
  for different interaction types.}
\label{tab:2comp_stabilitycond_N}
\begin{ruledtabular}
\begin{tabular}{c @{\qquad} c @{\qquad} c @{\qquad} c @{\qquad} c}
 interaction type & $C_0$ & $C_1$ & $C_{01}$ & $n$ \\
\hline
$a_0\le0,\,\tilde{a}_1\le0$ & -1.835 & -2.570 & 0.656  & 1  \\
$a_0\ge0,\,\tilde{a}_1<0$   & -1.378 & -2.570 & 1.360  & 1  \\
$a_0<0,\,\tilde{a}_1\ge0$   & -1.835 & -1.940 & 2.246  & 3  
\end{tabular}
\end{ruledtabular}
\end{table}

Before we discuss the structure of \eqref{eq:2comp_stabilitycond_kF}
we formulate an equivalent stability condition in terms of the number
of particles $N=N_1=N_2$ of each component. For given values of the
two scattering lengths the maximum local Fermi momentum and the
maximum chemical potential is calculated. From the solution of the
extremum condition \eqref{eq:2comp_extremumcond_2} for these
parameters the corresponding maximum particle number is
determined. This numerical data is fitted by a suitable
parameterization of the stability condition in terms of the particle
number and the two scattering lengths $a_0/\ell$ and
$\tilde{a}_1/\ell$
\eqmulti{ \label{eq:2comp_stabilitycond_N}
  &C_0 \Big(\!\sqrt[6]{N} \frac{a_0}{\ell}\Big) +
  C_1^3 \Big(\!\sqrt[6]{N} \frac{\tilde{a}_1}{\ell}\Big)^3\\ 
  &\qquad + C_{01}^{n+1} \Big(\!\sqrt[6]{N} \frac{a_0}{\ell}\Big)
  \Big(\!\sqrt[6]{N} \frac{\tilde{a}_1}{\ell}\Big)^n \le 1 .
}
This parameterization is constructed in analogy to the
single-component case \eqref{eq:1comp_stabilitycond_N}; the additional
cross-term is necessary to achieve a similar accuracy with typical
deviations below 1\%. The parameters $C_0$, $C_1$, and $C_{01}$ have
to be fitted for each combination of signs of the two scattering
length separately. The value of $n$ is not included in the fitting
procedure but chosen by hand. The resulting values are summarized in
Table \ref{tab:2comp_stabilitycond_N}.

Figure \ref{fig:2comp_stabmap} illustrates the dependence of the
maximum particle number resulting from this stability condition on the
s- and p-wave scattering lengths. The contour plot shows the 
logarithm of the maximum particle number for each component as
function of $a_0/\ell$ and $\tilde{a}_1/\ell$. The first gross observation is
that attractive s- and p-wave interactions with similar scattering
lengths set similar restrictions to the stability of the two-component
Fermi gas. For example, a pure s-wave interaction with
$a_0/\ell=-0.05$ leads to a maximum particle number of
$N_{\text{max}}\approx 1.7\times10^6$. In comparison a pure p-wave
interaction with same scattering length $\tilde{a}_1/\ell=-0.05$ causes a
collapse at even lower particle numbers of
$N_{\text{max}}\approx2.2\times10^5$. 

If both interaction parts are attractive they cooperate and cause an
instability at lower particle numbers or densities. 

If the interaction is attractive in one partial wave and repulsive in
the other than the repulsive part leads to a stabilization, i.e., it
increases the maximum particle number. Here a significant difference
between s- and p-wave interactions arises: The stabilization caused by
a repulsive s-wave interaction is rather weak. Compared to a pure
p-wave interaction with $\tilde{a}_1/\ell=-0.05$ the presence of a
s-wave repulsion of the same magnitude $a_0/\ell=0.05$ increases the
maximum particle number only from $2.2\times10^5$ to $8.9\times10^5$. In
the opposite case of an attractive s-wave interaction a p-wave
repulsion of same magnitude will always lead to an absolute
stabilization, i.e., there is no collapse for arbitrary large particle
number despite of the s-wave attraction. We will study these special
effects in detail in the following section.

These results clearly demonstrate that it is necessary to include
the p-wave interaction if the scattering length $\tilde{a}_1$ is roughly in
the same order of magnitude as the s-wave scattering length. Even if
the ratio of the scattering lengths, $\tilde{a}_1/a_0$, is approximately 0.3
dramatic effects like the p-wave stabilization, which is discussed in
the next section, can occur. As can be seen from Figure
\ref{fig:2comp_stabmap} the p-wave interaction may be neglected only
if the ratio $\tilde{a}_1/a_0$ is smaller than 0.1.

\subsection{Mean-Field Instability: p-Wave Stabilization}

\begin{figure}
\includegraphics[height=0.27\textheight]{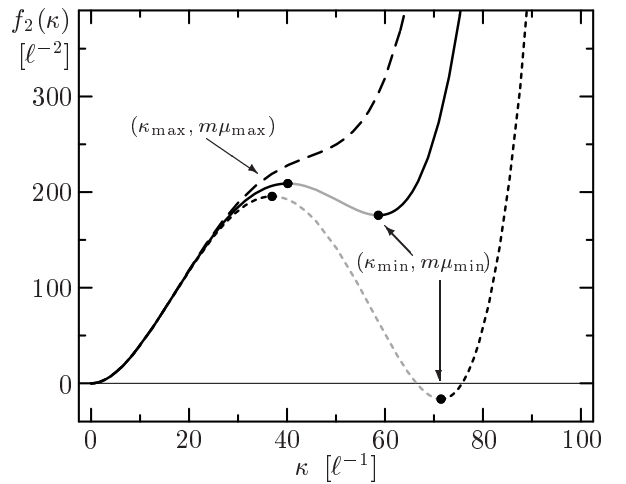}
\caption{Right hand side $f_2(\kappa)$ of the extremum condition
  \eqref{eq:2comp_extremumcond_2} as function of the Fermi momentum for
  attractive s-wave interaction with $a_0/\ell=-0.05$ and a repulsive
  p-wave interaction with $\tilde{a}_1/\ell=0.014$ (dotted curve), $0.015$
  (solid), and  $0.016$ (dashed). The gray segments of the curves
  correspond to maxima of the energy density.}
\label{fig:2comp_extremumcond_a0-a1+}
\end{figure}

Several new phenomena occur due to the competition between an
attractive s-wave ($a_0<0$) and a repulsive p-wave interaction
($\tilde{a}_1>0$). To understand the origin of these phenomena, which
are a unique property of this type of interactions, we investigate the
right hand side $f_2(\kappa)$ of the extremum condition
\eqref{eq:2comp_extremumcond_2}. Figure
\ref{fig:2comp_extremumcond_a0-a1+} depicts the dependence of
$f_2(\kappa)$ on the local Fermi momentum for a s-wave scattering
length $a_0/\ell=-0.05$ and three slightly different positive p-wave
scattering lengths in the range $a_0/\ell=0.014\dots0.016$.

Due to the dominant $\kappa^5$-dependence any repulsive p-wave
interaction causes $f_2(\kappa)$ to grow fast for large Fermi
momenta. Thus the maximum is only local and does not determine
necessarily the maximum Fermi momentum or chemical potential as in the
cases with attractive or vanishing p-wave interaction. If the p-wave
repulsion is sufficiently strong the local maximum vanishes completely
and $f_2(\kappa)$ is a monotonically growing function. Then a solution
of the extremum condition exists for any density, chemical potential
or particle number. An example is shown by the dashed curve in Figure
\ref{fig:2comp_extremumcond_a0-a1+}. It can be seen from equation
\eqref{eq:2comp_extremumcond_2} that the local maximum disappears if
the ratio of the two scattering lengths fulfills the condition
\eq{ \label{eq:2comp_absolutestab}
  \frac{\tilde{a}_1}{|a_0|} \ge \frac{2}{3\pi^{2/3}} \approx 0.311 .
}
If this condition is fulfilled the p-wave repulsion causes an
\emph{absolute stabilization} of the system against s-wave induced
collapse. In this case the total mean-field contribution of the
interactions is always repulsive and grows monotonically with density.

Notice that the p-wave scattering length necessary for this
stabilization is only approx. $1/3$ of the modulus of the s-wave
scattering length. Obviously the p-wave interaction may have drastic
influence on the stability even if it is significantly weaker than the
s-wave interaction in terms of scattering lengths.

For weaker p-wave repulsions $f_2(\kappa)$ still shows a local maximum
and in addition a (local) minimum at larger Fermi momenta. Examples
are shown by the solid and dotted curves in Figure
\ref{fig:2comp_extremumcond_a0-a1+}. In this case the extremum
condition has two separate branches that correspond to local minima of
the energy density, which are separated by branch of local maxima
(gray segments). The branch at lower Fermi momenta corresponds to the
usual family of low-density solutions that were obtained with other
types of interactions too. It ends up at the local maximum with
$\kappa_{\text{max}}$ given by \eqref{eq:2comp_stabilitycond_kFmax}
and $m\mu_{\text{max}} = f_2(\kappa_{\text{max}})$. The solution
branch at higher Fermi momenta gives rise to a new family of
high-density solutions, which are unique for this type of
interaction. It is bounded from below by the local minimum
$(\kappa_{\text{min}},\mu_{\text{min}})$ and raises up to arbitrary
Fermi momenta and chemical potentials.

\begin{figure}
\includegraphics[height=0.27\textheight]{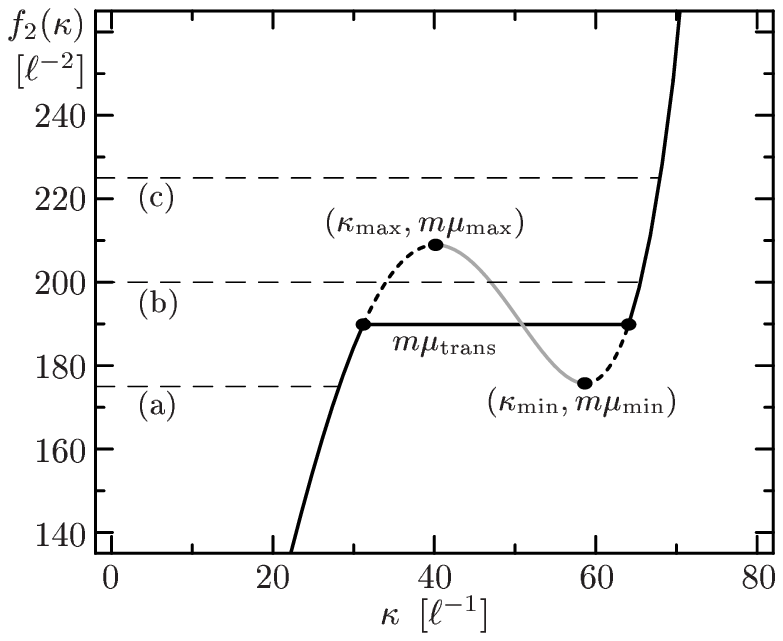}
\par\vspace{5pt}
\includegraphics[height=0.27\textheight]{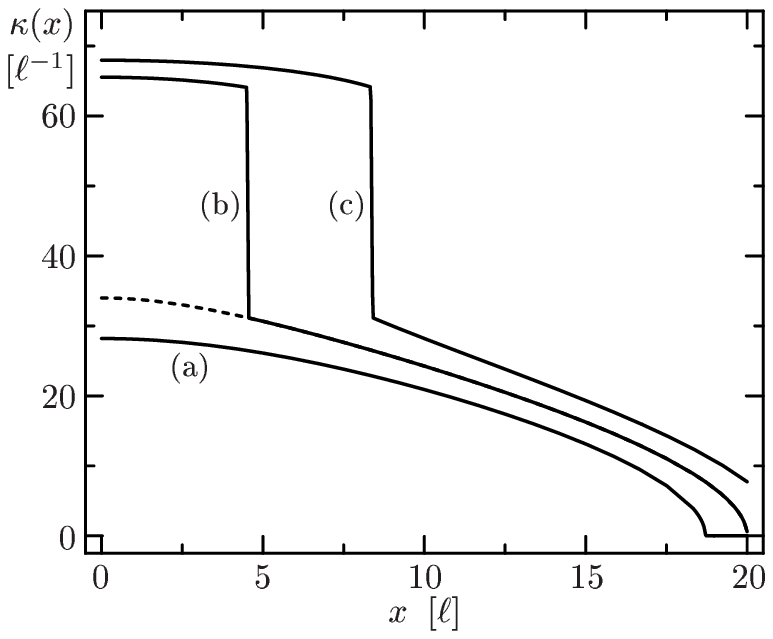}
\caption{Upper plot: $f_2(\kappa)$ of the extremum condition
  \eqref{eq:2comp_extremumcond_2} as function of the Fermi momentum for
  an interaction with $a_0/\ell=-0.05$ and $\tilde{a}_1/\ell=0.015$. Lower
  plot: Distribution of local Fermi momenta $\kappa(x)$ for a
  spherical trap of oscillator length $\ell$ for the three different
  chemical potentials marked in the upper panel. Solid curves show the
  equilibrium profiles, dotted curves show metastable configurations.}
\label{fig:2comp_highdens}
\end{figure}

For values of $[\mu-U(\xV)]$ between $\mu_{\text{min}}$ and
$\mu_{\text{max}}$ the extremum condition has two solutions
$\kappa_{\text{low}}$ and $\kappa_{\text{high}}$ with
$f_2(\kappa_{\text{low}}) = f_2(\kappa_{\text{high}})$, see Figure
\ref{fig:2comp_highdens}. In equilibrium the one with lower free
energy density \eqref{eq:2comp_energydens_f_1} is realized. We define
a chemical potential $\mu_{\text{trans}}$ at which the free energy
densities of both branches are equal; the value of
$\mu_{\text{trans}}$ can be determined numerically. Since we expect
the solution $\kappa(\xV)$ to correspond to a minimum of the energy
functional at each point $\xV$, for $[\mu-U(\xV)]<\mu_{\text{trans}}$
the low-density branch gives the equilibrium solution and for
$[\mu-U(\xV)]>\mu_{\text{trans}}$ the high-density branch does. This
gives rise to a Maxwell construction for the r.h.s. of the extremum
condition \eqref{eq:2comp_extremumcond_2} as illustrated in the upper
plot of Figure \ref{fig:2comp_highdens}. The dotted parts of the lower
and upper branch in Figure \ref{fig:2comp_highdens} correspond to
local minima and may occur as metastable states that eventually
undergo a transition to the energetically lower equilibrium solution.

The structure of the density distribution depends crucially on the
value of the chemical potential $\mu$, i.e., the particle number. The
upper plot of Figure \ref{fig:2comp_highdens} shows the r.h.s. of the
extremum condition \eqref{eq:2comp_extremumcond_2} for an interaction
with $a_0/\ell=-0.05$ and $\tilde{a}_1/\ell=0.015$. The dashed
horizontal lines mark three different chemical potentials. The lower
plot shows the radial dependencies of the local Fermi momenta
$\kappa(x)=(6\pi^2 \rho(x))^{1/3}$ for these three chemical potentials
assuming a spherical trap with oscillator length $\ell$.  For chemical
potentials $\mu<\mu_{\text{trans}}$ --- case (a) --- the equilibrium
solution is completely on the low-density branch and we obtain the
usual smooth density profile. If $\mu>\mu_{\text{trans}}$ --- case (b)
and (c) --- then the equilibrium solution in the center of the trap is
given by the high-density branch, while the solution for the outer
regions of the trap is given by the low-density branch. Thus the
equilibrium density profile shows a jump in density by typically one
order of magnitude as one approaches the center of the trap. The
location of the discontinuity is always given by the equation
$[\mu-U(\xV)]=\mu_{\text{trans}}$ which reflects the mechanical
equilibrium between the low and the high-density phase, i.e., the
equality of the pressures $p = -\FC[\kappa](\xV)$ at the boundary
between the two phases.

If the chemical potential is still below $\mu_{\text{max}}$ --- case
(b) --- then a solution with a smooth low-density profile all over the
trap may exist as metastable state. Due to density fluctuations this
state may undergo a transition to the energetically preferred
equilibrium state, which includes the high-density phase.

The physical origin of the high-density phase is quite intuitive: The
usual stability is determined by a competition between kinetic energy,
which favors low densities, and attractive mean-field, which prefers
higher densities. If the attractive mean-field becomes too strong,
than the kinetic energy is not able to stabilize the system anymore
and the mean-field collapse occurs. In case of the high-density phase
the attractive mean-field generated by the s-wave interaction has
already overcome the stabilizing effect of the kinetic
energy. Nevertheless the collapse is prevented by the repulsive p-wave
contribution which grows for higher densities faster than the s-wave
attraction and inhibits a further increase of density. We call this
new phenomenon \emph{p-wave stabilized high-density phase} --- in
contrast to the low-density phase stabilized by the kinetic energy,
which is still present in the peripheral regions of the trap.

The situation discussed so far assumes a repulsive p-wave interaction
that is slightly to weak to cause the absolute stabilization according
to \eqref{eq:2comp_absolutestab}. If the p-wave strength is decreased
further, then the values of $\mu_{\text{min}}$, $\mu_{\text{max}}$, and
$\mu_{\text{trans}}$ also decrease. If the ratio of the p-wave and
s-wave scattering lengths drops below the limit
\eq{
  \frac{\tilde{a}_1}{|a_0|} 
  < \sqrt[3]{\frac{160}{729\pi^2}} \approx 0.281
}
then the chemical potential of the minimum is negative,
$\mu_{\text{min}}<0$. An example is shown by the dotted curve in
Figure \ref{fig:2comp_extremumcond_a0-a1+}. For even weaker p-wave
interactions with 
\eq{ \label{eq:2comp_nohighdens} 
  \frac{\tilde{a}_1}{|a_0|} < 0.274 
}
$\mu$ can be negative. That means that the high-density solution forms
a self-bound state independent of the trapping potential. Therefore as
soon as the maximum chemical potential $\mu_{\text{max}}$ is exceeded
the gas collapses into a self-bound high-density state which is
independent of the trap.

We summarize the variety of structures that appear for interactions
with attractive s-wave ($a_0<0$) and repulsive p-wave part ($\tilde{a}_1>0$)
in the following list
\begin{description}
\item[$0.311<\tilde{a}_1/|a_0|$:] the p-wave repulsion stabilizes the system
  for arbitrary densities and particle numbers with a smooth
  low-density profile. For ratios of the scattering lengths 
  near the limit a smooth but significant increase of the central 
  density occurs.
\item[$0.274 < \tilde{a}_1/|a_0| < 0.311$:] For chemical potentials below
  $\mu_{\text{max}}$ or particle numbers below the corresponding maximum
  particle number \eqref{eq:2comp_stabilitycond_N} the usual low
  density solution exists. Above these values the p-wave stabilized
  high-density phase appears, i.e., the density in the central region
  of the trap is increased by typically one order of magnitude
  compared to the low-density profile in the outer regions. 
\item[$\tilde{a}_1/|a_0| < 0.274$:] Below $\mu_{\text{max}}$ or
  $N_{\text{max}}$ a stable solution with the regular low-density
  profile exists. Above the system collapses; the high-density phase is
  not stable anymore.
\end{description}
This subtle dependence on the ratio of the scattering lengths is
illustrated in Figure \ref{fig:2comp_stabmap}. The white region for
``strong'' repulsive p-wave interactions shows the domain of absolute
stabilization. The solid line corresponds to the condition
\eqref{eq:2comp_absolutestab} for absolute stabilization, the dashed
line to condition \eqref{eq:2comp_nohighdens} for the stability of the
high-density phase. The small area between those lines represents the
parameter region where the p-wave stabilized high-density phase exists
if the maximum particle number is exceeded.
 
\begin{figure}
\includegraphics[width=0.9\columnwidth]{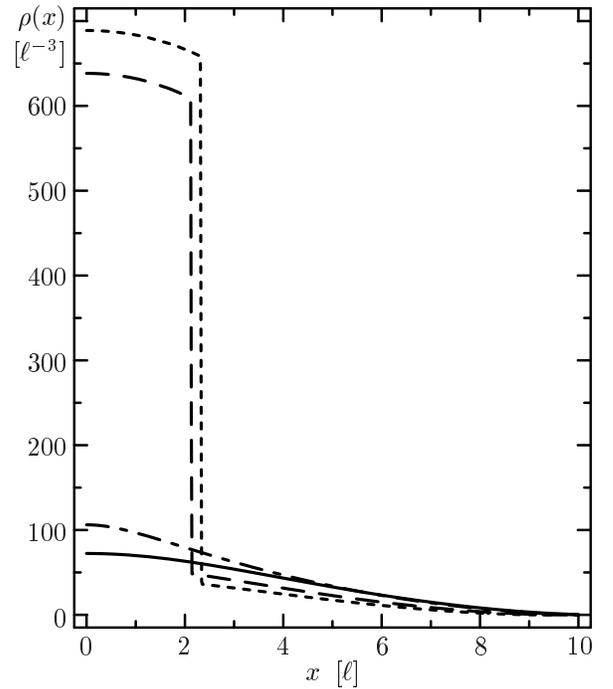}
\caption{Evolution of the density profile of a two-component system of
  $N=N_1=N_2=60000$ particles with p-wave scattering length
  $\tilde{a}_1/\ell=0.03$ according to the \emph{Gedankenexperiment} described
  in the text. The s-wave scattering length is tuned in the range
  $a_0/\ell=-0.095$ (solid), $-0.1$ (dash-dotted), $-0.101$ (dashed), and
  $-0.102$ (dotted).  }
\label{fig:2comp_gendankenexp}
\end{figure}

To conclude this section we perform a \emph{Gedankenexperiment}:
Assume a two-component Fermi gas of $N=N_1=N_2=60000$ particles in
each component trapped in a spherical oscillator potential with
$\ell=1\mu m$. May the interaction be composed of a repulsive p-wave
part with $\tilde{a}_1/\ell=0.03$ and an attractive s-wave component
which can be tuned within a small range
$a_0/\ell=-0.095,\dots,-0.102$, e.g., by using a Feshbach
resonance. Figure \ref{fig:2comp_gendankenexp} shows the evolution of
the density profile of the Fermi gas if the strength of the attractive
s-wave interaction is increased slowly such that density fluctuations
are negligible. For $a_0/\ell=-0.095$ (solid curve) and $-0.1$
(dash-dotted) we observe a smooth low density profile, where the
central density increases slightly with increasing s-wave
attraction. A dramatic change happens if the attraction is increased
to $a_0/\ell=-0.101$ (dashed). For this interaction strength the
particle number of the system is already above the maximum particle
number given by \eqref{eq:2comp_stabilitycond_N} and the high-density
phase appears and occupies a rather large volume. Further increase of
the s-wave attraction (dotted) causes a growth of the high-density
phase.  If the limit $a_0/\ell=-0.11$ is reached than part of the
high-density component is self-bound and the system is expected to
collapse.

\section{Summary and Conclusions}
                                                                                    
We formulated a simple and transparent model to describe the structure
and stability of degenerate multi-component Fermi gases trapped in an
external potential. In a first step we derived an Effective Contact
Interaction (ECI) for all partial waves that reproduces the exact
two-body energy spectrum when used in a mean-field model
space. Including the s- and p-wave parts of the ECI we constructed the
energy density of the inhomogenous Fermi gas in a mean-field
calculation using the Thomas-Fermi approximation. By functional
minimization of the energy we obtained a set of coupled polynomial
equations for the ground state density profile of the system.  We
showed that the combination of s- and p-wave interactions leads to a
rich variety of phenomena in trapped degenerate Fermi gases.

In the single-component system the p-wave part is the leading
interaction term since s-wave scatterings are prohibited by the Pauli
principle. Attractive p-wave interactions cause a mean-field
instability of the one-component gas if a certain maximum density is
exceeded. We derived explicit stability conditions in terms of
the density or particle number and the p-wave scattering length.

The interplay between s- and p-wave interactions leads to several new
effects in the two-component Fermi gas. We discussed the dependence of
the mean-field instability on the s- and p-wave scattering lengths and
derived also for this case closed stability conditions. It turns out
that attractive s-wave as well as attractive p-wave interactions can
cause a mean-field collapse. In addition a repulsive interaction part
leads to stabilization, i.e., an increase of the maximum possible
density of the Fermi gas.

Interactions with attractive s-wave and repulsive p-wave parts show
several special properties. If the p-wave scattering length exceeds
about $1/3$ of the modulus of the s-wave scattering length the s-wave
attraction is fully compensated and no mean-field collapse occurs
anymore at high densities. In the transition region towards this
absolute stabilization a distinct high-density phase may appear in the
center of the trap, which is stabilized by the p-wave repulsion alone.
  
We conclude that the p-wave interaction may have an important
influence on the structure and stability of dilute degenerate Fermi
gases. Considering the simultaneous s- and p-wave Feshbach resonances
predicted for the ${}^{40}$K system \cite{Bohn00} it can be foreseen
that large values of p-wave scattering length will be available
experimentally. In a two-component ${}^{40}$K gas this Feshbach
resonance would allow to probe nearly the whole stability map shown in
Figure \ref{fig:2comp_stabmap} by modifying the magnetic field.
Alternatively, tightly confining optical traps \cite{ViGi01} generate
large values of the ratio of scattering length and oscillator length,
$a_l/\ell$, such that instabilities occur at much lower densities and
particle numbers ($N_{\text{max}}\propto \ell^6$; see
equations \eqref{eq:1comp_stabilitycond_N} and
\eqref{eq:2comp_stabilitycond_N}).

Concerning the envisioned observation of Cooper pairing in trapped
dilute Fermi gases two-component systems with strong attractive s-wave
interactions are favored \cite{DePa01,HaGe00}. Here the mean-field
instability limits the density of the normal Fermi gas. With a
suitably chosen repulsive p-wave interaction one could use the effect
of absolute stabilization, which we discussed, to allow higher
densities of the normal Fermi gas and thus increase the transition
temperature to a superfluid state \cite{HoFe97}.



\end{document}